\documentclass[%
 reprint,prd,
superscriptaddress,
nofootinbib,
 amsmath,amssymb,
 aps,
]{revtex4-2}
\usepackage{multirow}
\usepackage{wrapfig}
\usepackage{caption}
\usepackage{subcaption}
\usepackage{sidecap}
\usepackage{float}
\usepackage{bbold}

\usepackage{graphicx}
\usepackage{dcolumn}
\usepackage{bm}
\usepackage{xspace}
\xspaceaddexceptions{]\}}
\usepackage{hyperref}
\usepackage{ulem} 

\usepackage[dvipsnames]{xcolor}

\def\code#1{\texttt{#1}} 
\usepackage{todonotes}

\renewcommand{\vec}[1]{{\mathbf{#1}}}

\DeclareMathOperator{\Tr}{tr}
\newcommand{\model}{\ensuremath{\vec{m}}\xspace}
\newcommand{\dat}{\ensuremath{\vec{d}}\xspace}
\newcommand{\param}{\ensuremath{\vec{p}}\xspace}
\newcommand{\cov}{\ensuremath{\mathbf{\Sigma}}\xspace}
\newcommand{\invcov}{\ensuremath{\mathbf{\Sigma}^{-1}}\xspace}
\newcommand{\like}{\ensuremath{\mathcal{L}}\xspace}
\newcommand{\loglike}{\ensuremath{\ln{\mathcal{L}}}\xspace}

\newcommand{\photoz}{photo-$z$\xspace}

\newcommand{\lcdm}{\ensuremath{\Lambda\text{CDM}}\xspace}

\newcommand{\maglim}{MagLim\xspace}

\newcommand{\sect}[1]{Sec.~\ref{#1}\xspace}
\newcommand{\sects}[1]{Secs.~\ref{#1}\xspace}
\newcommand{\app}[1]{Appendix~\ref{#1}\xspace}

\newcommand{\fig}[1]{Fig.~\ref{#1}\xspace}

\newcommand{\eq}[1]{Eq.~(\ref{#1})\xspace}
\newcommand{\eqs}[1]{Eqs.~(\ref{#1})\xspace}
\newcommand{\tab}[1]{Table~\ref{#1}\xspace}

\def\hmath$#1${\texorpdfstring{{\rmfamily\textit{#1}}}{#1}}

\begin{document}

\preprint{APS/123-QED}

\title{Improving Photometric Galaxy Clustering Constraints With Cross-Bin Correlations}

\author{Jordan Krywonos}
  \email{jkrywonos@perimeterinstitute.ca }
  \affiliation{%
 Perimeter Institute for Theoretical Physics, 31 Caroline St N, Waterloo, ON N2L 2Y5, Canada
}%
\affiliation{%
 Department of Physics and Astronomy, York University, Toronto, ON M3J 1P3, Canada
}
\author{Jessica Muir}%
  \email{jmuir@perimeterinstitute.ca }
  \affiliation{%
 Perimeter Institute for Theoretical Physics, 31 Caroline St N, Waterloo, ON N2L 2Y5, Canada
}%
 \author{Matthew C. Johnson}%
   \email{mjohnson@perimeterinstitute.ca}
\affiliation{%
 Perimeter Institute for Theoretical Physics, 31 Caroline St N, Waterloo, ON N2L 2Y5, Canada
}%
\affiliation{%
 Department of Physics and Astronomy, York University, Toronto, ON M3J 1P3, Canada
}

\date{\today}

\begin{abstract}
Clustering studies in current photometric galaxy surveys focus solely on auto-correlations, neglecting cross-correlations between redshift bins. We evaluate the potential advantages and drawbacks of incorporating cross-bin correlations in Fisher forecasts for the Dark Energy Survey (DES) and the forthcoming Rubin Observatory Legacy Survey of Space and Time (LSST). Our analysis considers the impact of including redshift space distortions (RSD) and magnification in model predictions, as well as systematic uncertainties in photometric redshift distributions (\photoz). While auto-correlations alone suffer from a degeneracy between the amplitude of matter fluctuations ($\sigma_8$) and galaxy bias parameters, accounting for RSD and magnification in cross-correlations helps break this degeneracy - although more weakly than the degeneracy breaking expected from a combined analysis with other observables. Incorporating cross-bin correlations does not significantly increase sensitivity to \photoz systematics, addressing previous concerns, and self-calibrates \photoz systematics, reducing errors on \photoz nuisance parameters. We suggest that the benefits of including cross-correlations in future photometric galaxy clustering analyses outweigh the risks, but caution that careful evaluation is necessary as more realistic pictures of surveys' precision and systematic error budgets develop.
\end{abstract}

\maketitle


\section{Introduction}\label{sec:intro}

Characterizing the nature of dark energy, dark matter, and other physics affecting the large-scale Universe  has motivated significant investment in experiments mapping  cosmological large scale structure. 
Imaging, or photometric, galaxy surveys  represent an important class of these efforts, for which galaxy redshifts are estimated based on their brightness in different color filters.  Examples include the Dark Energy Survey (DES)\footnote{\url{http://www.darkenergysurvey.org/}}
~\cite{flaugher15,DES:Y3}, which completed six years of observations in 2019 and for which the final cosmology analysis is in progress, as well as next-generation surveys like the Vera Rubin Observatory Legacy Survey of Space and Time (LSST),\footnote{\url{http://www.lsst.org/}} which is set to begin science operations in the near future. 
While imaging surveys are qualitatively unique in their ability to probe weak gravitational lensing distortions to galaxy shapes, they additionally produce rich data for measuring galaxy clustering. Photometric clustering measurements benefit from  wide survey area, complete overlap with weak lensing shear measurements, and high galaxy number density. Three-dimensional information can be obtained by measuring the projected angular correlations of galaxies split into tomographic bins.  

DES is the the only Stage III imaging survey that has included photometric clustering in its main cosmology analyses, as part of a so-called $3\times 2$pt analysis combining clustering with measurements of galaxy-galaxy lensing and cosmic shear~\cite{DES:Y1cosm,DES:Y3}. In contrast, combined clustering and lensing analyses from the Kilo-Degree Survey (KiDS)\footnote{\url{https://kids.strw.leidenuniv.nl/}} and the Subaru Hyper Suprime-Cam (HSC) survey\footnote{\url{https://hsc.mtk.nao.ac.jp/ssp/}}, have mainly used imaging survey data for cosmic shear measurements, and measured galaxy clustering using spectroscopic catalogs~\cite{Heymans:2020gsg,Dvornik:2022xap,More:2023knf,Sugiyama:2023fzm,Miyatake:2023njf}. Exceptions to this include Refs.~\cite{Vakili:2020dwl} and~\cite{Nicola:HSCclustering} which use tomographic clustering of galaxies measured with KiDS and HSC, respectively, to constrain galaxy sample properties at fixed cosmology, and Ref.~\cite{Xu:2023zlv}'s  proof-of-concept photometric clustering analysis of  DESI\footnote{\url{https://www.desi.lbl.gov/}} imaging data.  Past examples of photometric clustering analyses have been done with data from CFHTLens~\cite{Coupon:2011mz}, SDSS~\cite{Padmanabhan:2006egz,Blake:2006kv,Estrada:2008em,dePutter:2012sh,Ho:2013lda,DES:2015vbk}, 2MASS~\cite{Balaguera-Antolinez:2017dpm}, and NVSS~\cite{Blake:2001bg}. Also, while our focus is on analyses of the full shape of  angular correlation functions, we note that a number of studies have extracted cosmological information from the study of the baryon acoustic oscillation feature (BAO) in photometric galaxy samples~\cite{desy3bao,Seo:2012xy,Crocce:2011mj,Carnero:2011pu,Huetsi:2009zq}. Looking ahead, photometric clustering is included as a key observable in Stage IV survey forecasts from the LSST Dark Energy Science collaboration (DESC)\footnote{\url{http://lsstdesc.org/}}~\cite{LSST:DESC_SRD}, Euclid~\cite{EUCLID:definitionreport,Euclid:photoclustering_opt}, and the Nancy Grace Roman Space Telescope~\cite{Spergel:2015sza,Eifler:2020vvg}.

A key choice in the design of a photometric galaxy clustering study is whether to include cross-correlations between galaxies in different redshift bins. The summary statistics used in DES analyses of the survey's first year (Y1)~\cite{DES:Y1cosm} and first three years (Y3)~\cite{DES:Y3,DES:maglim2x2pt_results,DES:redmagic2x2pt_results} of data include correlations between galaxy positions only for pairs  in the same tomographic bin.  This choice was motivated by the fact that forecasts showed cross-bin correlations to have minimal impact on the precision of cosmology constraints, combined with a concern that they introduce greater sensitivity to  systematic errors, particularly related to photometric redshifts~\cite{DES:maglim_optimization}. As survey precision and methodology improve, it is worth reconsidering this choice. Promisingly, several studies have indicated the potential for  cross-bin correlations to self-calibrate astrophysical and observational uncertainties, including a DES study of the impact of magnification on galaxy clustering~\cite{DES:magnification}, an LSST forecast examining redshift evolution of galaxy bias and magnification  within tomographic bins~\cite{Pandey:2023tjn}, and approaches to constraining photometric redshift uncertainties explored in Ref.~\cite{Nicola:HSCclustering}'s  HSC analysis and Ref.~\cite{Schaan:2020qox}'s LSST forecasts. Outside of these studies, choices in photometric clustering forecasts are mixed: the LSST Dark Energy Science Collaboration Science Requirements (DESC SRD) document~\cite{LSST:DESC_SRD} considers auto-correlations only, as do many studies based on that document (e.g. Refs.~\cite{Fang:2023efj,LSSTPratZuntz:2022sql}) and forecasts for  Roman~\cite{Eifler:2020vvg}, while Euclid photometric clustering forecasts~\cite{Euclid:forecast_validation,Euclid:photoclustering_opt,Euclid:rsd,TanidisCamera:rsd:2019teo} and some LSST forecasts (e.g. Refs.~\cite{Lorenz:2017iez,Mahony:magnification}) include all cross-bin correlations. 

Ultimately, when deciding whether or not to include cross-bin correlations in a clustering analysis, we must weigh potential gains in constraining power against robustness to biases from systematics. The goal of this paper is to explore how this balance is influenced by two considerations which have become increasingly relevant with advancing survey precision: 
\begin{enumerate}
\item Contributions to observed galaxy clustering from redshift space distortions (RSD) and magnification. 
\item Uncertainties in the shape of galaxy samples' photometric redshift distributions.
\end{enumerate}

RSD and magnification are interesting in this context because they enhance the cosmological signal in cross-bin correlations, and because their inclusion in model predictions has only recently been necessitated by surveys' accuracy requirements. 
RSD arise from the fact that peculiar motions of galaxies  change the observed redshift for a galaxy relative to that from Hubble expansion alone. This affects projected galaxy clustering at large angular scales, potentially moving galaxies between neighboring redshift bins, with the size of the effect controlled by the growth rate of large-scale structure. Magnification refers to contributions to observed clustering from variations in a survey's limiting magnitude caused by weak gravitational lensing. Both effects were determined to be negligible for DES Y1 modeling~\cite{DES:Y1methods,DES:Y1cosm}, but they were added to DES Y3 galaxy clustering calculations  after simulated analyses  showed  this was needed to avoid biased cosmological inferences at the increased precision~\cite{DES:Y3methods,Fang:2020,DES:magnification}. The novelty of this accuracy requirement means that RSD and magnification  have not always been included in photometric clustering forecasts, including the study that guided the choice to leave cross-bin correlations out of the DES Y3 analysis ~\cite{DES:maglim_optimization}. 

Redshift distribution uncertainties are relevant because they are the systematic effect that is most likely to limit our ability to make accurate inferences using cross-bin correlations. Photometric clustering inferences generally incorporate photometric redshift (hereafter, \photoz) uncertainties in two ways. First, the shape of the redshift distributions $n_i(z)$ for each bin $i$ have overlapping tails rather than sharp divisions between bins, reflecting the fact that uncertainties in \photoz estimates cause samples selected based on \photoz  to have extended true redshift distributions.  Second, residual uncertainties in these fiducial distributions are accounted for by marginalizing over nuisance parameters  which change the mean (shift) or width (stretch) of each $n_i(z)$ distribution. Though  \photoz uncertainty models with greater flexibility have been proposed, as in e.g. Refs.~\cite{DES:hyperrank,Schaan:2020qox,Hadzhiyska:2020xob}, for DES Y3 (which uses auto-correlations only) this shift-stretch treatment was shown to be sufficient to obtain unbiased cosmology constraints~\cite{DES:lenswz,DES:2022redshiftcal,DES:maglim2x2pt_results,DES:redmagic2x2pt_results}.  In practice this method is what is used for most future-survey forecasts. Roughly speaking, the amount of correlation between redshift bins $i$ and $j$ depends on the product of their redshift distributions, $n_i(z)\,n_j(z)$. Especially for bins with large separations, features in the tails of those distributions can have a dramatic impact on this product, and thus on the correlation between the galaxy samples. Given this, when analysing cross-bin correlations, it is likely that marginalizing over variations in the mean and width of the redshift bins will no longer be sufficient to capture the impact of \photoz uncertainties on the observables. 

Here we use Fisher forecasts for DES Y3 and LSST Y1 to quantify the benefits of and risks of adding cross-bin correlations to a photometric clustering analysis. 
For simplicity, we focus on an analysis of galaxy clustering alone, investigating  constraints on $\Omega_{\rm c}$ and $\sigma_8$ while fixing other cosmological parameters. This can be viewed as an optimistic representation of clustering's constraining power in a scenario where the fixed parameters are subject to tight priors based on other measurements. 
Our investigation proceeds as follows. We begin in \sect{sec:modeling} by describing our methods for modeling photometric galaxy clustering and then for Fisher forecasting in \sect{sec:forecastmethods}. \sect{sec:rsdmag} characterizes how RSD and magnification influence the information content of the cross-bins, first by evaluating the significance of the effects' contributions to observables, then by forecasting their impact on the precision and accuracy of cosmological inference. We next turn in \sect{sec:photoz} to study the impact of variations in galaxy redshift distributions, showing that adding cross-bin correlations lessens standard nuisance parameters' ability to protect cosmological inference from biases due to \photoz uncertainties. As part of that investigation we propose a computationally inexpensive procedure for correcting model predictions from biases due to noise in $n(z)$ distributions. Finally in \sect{sec:conclusion} we summarize our results and discuss possible approaches for addressing this challenge in future analyses.

\section{Modeling galaxy clustering}\label{sec:modeling}

We consider galaxy clustering, the correlations between galaxy number density fields, quantified using angular power spectra. 
For two fields $A$ and $B$ describing fluctuations on the sky, their angular cross-power spectum $C^{AB}_{\ell}$ is defined as the variance of their spherical harmonic components $a_{\ell m}$,
\begin{equation}\label{eq:fishercovariance}
    \langle a_{\ell m}^A(a^B_{\ell'm'})^*\rangle=\delta_{\ell\ell'}\delta_{mm'}C_\ell^{AB}, 
\end{equation}
where the angled brackets $\langle\dots\rangle$ signify an average over many realizations of density fluctuations in the Universe. However, we can actually only observe one realization of a cosmological field, so in practice we use the estimator of the observed power spectrum
\begin{equation}
    \hat{C}_\ell^{AB} = \frac{1}{2\ell+1}\sum_{m=-\ell}^{\ell}|(a_{\ell m}^A)^*a_{\ell m}^B|
\end{equation}
such that \(\langle \hat{C_\ell}\rangle = C_\ell\). 
Assuming Gaussian fluctuations, the measured spectrum $\hat{C}_{\ell}^{AB}$ has covariance
\begin{align}\label{eq:cov}
    &\rm{Cov}(\hat{C}^{AB}_{\ell},\hat{C}^{DE}_{\ell'}) = \left\langle\left(\hat{C}^{AB}_{\ell}-C^{AB}_\ell\right)\left(\hat{C}^{DE}_{\ell'}-C^{DE}_{\ell'}\right)\right\rangle \\&\quad\quad= \frac{\delta_{\ell\ell'}}{f_{\rm sky}(2\ell+1)}\left[C^{AD}_\ell C^{BE}_\ell+C^{AE}_\ell C^{BD}_\ell\right].\notag
\end{align}
Here, $f_{\rm sky}$ is the fraction of sky area covered by our measurements, $\delta_{\ell \ell'}$ is a Kronecker delta, and $A$, $B$, $D$, $E$ label tracer fields. 

The power spectra include contributions 
\begin{equation}
   C_{\ell}^{AB} \equiv  \tilde{C}_{\ell}^{AB} + \delta_{AB}\bar{n}_A^{-1},
\end{equation}
where the first term $\tilde{C}_{\ell}^{AB}$ is the model prediction for clustering due to underlying density fluctuations and  the second term describes shot noise contributions to auto-power spectra. The parameter $\bar{n}_A$ is the number density of galaxies per steradian.

We model the correlation between two galaxy samples, which e.g. could correspond to tomographic bins, as 
\begin{equation}
    \tilde{C}^{AB}_{\ell} = \frac{2}{\pi} \int \frac{dk}{k} k^3 P_\Phi(k) \Delta_A(\ell,k) \Delta_B(\ell,k)\label{eq:cellint}
\end{equation}
 where $P_\Phi(k)$ is the power spectrum of primordial curvature perturbations and the tracers' transfer functions are $\Delta_X(\ell,k)$, where $X$ labels the galaxy sample. 
The galaxy clustering transfer function receives several contributions,  
\begin{equation}
    \Delta_X  = \Delta_X^{D} + \Delta_X^{R} + \Delta_X^{M}. 
\end{equation}
Here $\Delta_X^D$ is the largest contribution, encapsulating the fact that galaxies trace the underlying total matter density field, while $\Delta_X^R$ (RSD) and $\Delta_X^M$ (magnification) are the next most important contributions~\cite{Lorenz:2017iez}.
For this work, we model these contributions using a linear galaxy bias and linear RSD~\cite{Kaiser1987}, which give transfer functions of the form~\cite{Fang:2020}
\begin{align}
    \Delta_X^{D}(\ell,k) &= b^X_g \int dz\, n_X(z) T_\delta(k,z) j_\ell(k \chi(z)) \label{eq:DeltaD}\\
    \Delta_X^R(\ell,k) &= -\int dz\, f(z) n_X(z) T_\delta(k,z) j_\ell''(k \chi(z)) \label{eq:DeltaR}\\
    \Delta_X^M(\ell,k) &= \frac{3\ell(\ell+1)H_0^2 \Omega_{\rm m}}{2} \label{eq:DeltaM}\\&\quad\times\int dz\, \frac{(1+z)}{cH(z)k^2} W^M_X(z) T_\delta(k,z) j_\ell(k \chi(z))\notag
\end{align}
Here $\chi$ is the comoving distance, $T_\delta(k,z)$ is the transfer function of matter density perturbations, $f(z)$ is linear growth rate, and the derivative of the spherical Bessel function $j_\ell(x)$ is defined as
\begin{equation}
    j_\ell'(x)=\frac{\partial j_\ell(x)}{\partial x}.
\end{equation}
Additionally, $b_g^X$ is the linear bias for galaxy sample $X$, while $n_X(z)$ is its normalized redshift distribution,
\begin{equation}
    n_X(z) = \frac{d\tilde{n}_X}{dz}\left[\int dz \frac{d\tilde{n}_X}{dz}\right]^{-1}.
\end{equation}
Here $\tilde{n}_X$ is the sample's average density over the sky. The function $W_X^M(z)$ is the lensing magnification window function, 
\begin{equation}
    W_X^M(z) = b^X_m \int_z^\infty dz'\, n_X(z') \frac{\chi(z') - \chi(z)}{\chi(z') \chi(z)}
\end{equation}
which depends on the sample's magnification bias $b^X_m$. More explicitly, for a magnitude limited survey, the magnification bias is defined as 
\begin{equation}
    b^X_m(z) = -2 + 5\frac{\partial \mathrm{log}_{10} \bar N(z,m< m_*)}{\partial m_*}
\end{equation}
with $\bar N(z,m< m_*)$ being the cumulative number of sources with magnitudes less than the magnitude cut of $m_*$.

Uncertainties in the redshift distributions are modeled by varying two parameters for each galaxy sample, a shift $\Delta z$ in the  mean of the redshift bin, and a stretch $\sigma_z$ which scales the width. If $n_0(z)$ is the functional form of the fiducial (unshifted, unstretched)  distribution and $\bar{z}_0$ is its mean redshift, the sample's redshift distribution is
\begin{equation}
    n(z) = \frac{1}{\sigma_z}n_0\left(\frac{z - \bar{z}_0 - \Delta z}{\sigma_z} + \bar{z}_0\right).
\end{equation}
As we discuss in more detail in \sect{sec:photoz}, information about the accuracy of the photometric redshift estimation and calibration procedure is incorporated via Gaussian priors on these shift and stretch parameters. 

We compute the power spectra using \code{CAMB}~\cite{CAMB,Lewis:1999bs,Howlett:2012mh,Challinor:2011bk} assuming the cosmological parameters listed in Table~(\ref{table:fidcosmoparams}). We compute the matter power spectrum using linear theory, though we do verify that  including nonlinear corrections does not substantially change our conclusions. Since we are concerned with effects impacting large scales, we perform the full triple-integral calculation described above rather than using a Limber approximation. Because we consider redshift distributions that are not smooth, we set the \code{CAMB} accuracy parameter \code{TimeStepBoost} to 15 in order to avoid inaccuracies in the $C_\ell$ calculations. 

We perform forecasts for two survey configurations, one matching the properties of DES Y3 and one representing the first year analysis of data from the Rubin Observatory (LSST Y1). This will allow us to situate our findings relative to a current state-of-the-art photometric clustering analysis while also exploring how  the increased precision of next-generation surveys impact our assessments. For both surveys, the observables we consider are the angular power spectra for multipoles $\ell> 20$. 
The value $\ell_{\rm min}=20$ is selected to avoid observational systematics impacting larger scale modes, though we also perform some forecasts with lower $\ell_{\rm min}=2$ to show how this changes our findings. In each redshift bin $i$, we impose scale cuts to protect against non-linear modeling uncertainties by removing  multipoles above $\ell_{\rm max}^i = k_{\rm max} \chi(\bar{z}^i)$, where $\bar{z}^i$ is the bin's mean redshift and $k_{\rm max}=0.3h{\rm Mpc}^{-1}$. 
Note that the linear matter power spectrum receives modest non-linear corrections at these multipoles, but for the purposes of our Fisher forecasts neglecting these effects does not impact our conclusions. Survey-specific modeling choices are described below.

\subsection{DES Y3}\label{sec:des}

We model our DES forecast on the properties of  the fiducial \maglim  galaxy sample used for galaxy clustering measurements in DES Y3 cosmology analyses~\cite{DES:Y3,DES:maglim2x2pt_results}. It contains 10.7 million galaxies selected using a redshift-dependent magnitude cut optimized for DES's science goals~\cite{DES:maglim_optimization}. We follow the DES Y3 analysis in focusing  our clustering analysis on  galaxies in four tomographic bins  with nominal edges at $z=[0.20,0.40,0.55,0.70,0.85]$. Using the survey area 4,143 ${\rm deg}^2$, we assign an equal  galaxy number density of $\bar{n} = 0.12\,\text{arcmin}^{-2} = 1.4\times 10^6{\rm sr}^{-1}$  to each bin. Fiducial values for linear galaxy bias, $b_g^i = [1.5, 1.8, 1.8, 1.9]$ (in order from lowest to highest redshift), are chosen to match those used in DES Y3 simulated analyses~\cite{DES:Y3methods}, and following DES Y3~\cite{DES:Y3,DES:magnification} we fix magnification biases to $b^i_m = [0.43, 0.30, 1.75, 1.94]$.  These quantities are summarized in \tab{table:fidparams}. 

\begin{figure}
    \begin{subfigure}[b]{\linewidth}
         \centering
         \includegraphics[width=\linewidth]{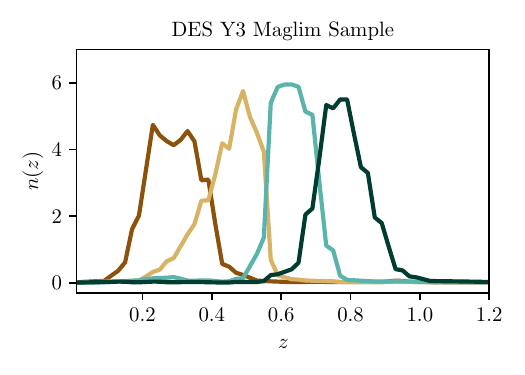}
         \label{fig:DESnz}
     \end{subfigure}
     \hfill
     \begin{subfigure}[b]{\linewidth}
         \centering
         \includegraphics[width=\linewidth]{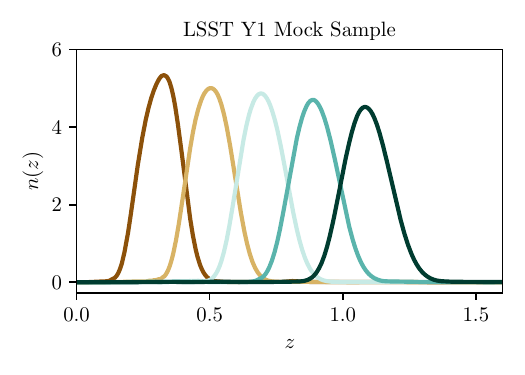}
         \label{fig:LSSTnz}
     \end{subfigure}
     \captionsetup{justification=raggedright}
    \caption{The DES (\textit{top panel}) and LSST (\textit{bottom panel}) lens sample redshift distributions for each redshift bin.}
    \label{fig:nzs}
\end{figure}

We use redshift distributions based on the \maglim redshift calibration study in Ref.~\cite{DES:2022redshiftcal}.  In the main DES Y3 analysis, the \maglim $n(z)$ distributions were inferred from Directional Neighborhood Fitting (DNF)~\cite{DeVicente:2015kyp,DES:2020aks} \photoz estimates, and calibrated using cross-correlations with spectroscopic galaxy samples~\cite{DES:lenswz}. Ref.~\cite{DES:2022redshiftcal} updated this calibration by using a self organizing map \photoz (SOMPZ) method,  which generates an ensemble of 6000 $n(z)$ realizations based on measured galaxy colors, also calibrated using spectroscopic cross-correlations. In \sect{sec:photoz} we will use this ensemble to test sensitivity to variations in $n(z)$ shapes. We set our fiducial redshift distribution equal to the average of the ensemble, shown in \fig{fig:nzs}, and adopt Ref.~\cite{DES:2022redshiftcal}'s Gaussian priors on shift and stretch parameters, listed in \tab{table:fidparams}. Those priors are centered  shift $\Delta z^i=0$ and stretch $\sigma_z^i=1$ for  all bins $i$, and their widths are equal to the the standard deviation of the ensemble of mean redshifts and bin widths\footnote{Bin width is measured as the difference between the redshifts marking the 16\% and 84\% quantiles of $n(z)$. This width measurement is preferred over computing standard deviation  $[\int dz \, n(z) (z-\bar{z})^2]^{1/2}$ because it is less sensitive to outliers from high-$z$ features in $n(z)$. }.

\subsection{LSST Y1}\label{sec:lsst}

We base our forecast for LSST Y1 on the DES SRD specifications for the LSST Y1 lens sample~\cite{LSST:DESC_SRD}, as well as implementations of those properties in Refs.~\cite{Zhang:2022,Fang:2020}. The survey area is 12,300 ${\rm deg}^2$, giving $f_{\rm sky}=0.3075$. Assuming a limiting magnitude of $i_{\rm lim}=24.1$ results in a total galaxy number density of 18 ${\rm arcmin}^{-2}$, which we divide evenly between five redshift bins with nominal edges at $z=[0.2, 0.4, 0.6, 0.8, 1.0, 1.2]$. Fiducial values for galaxy bias are $b_g^i=1.05/D(\bar{z}^i) = [1.56, 1.73, 1.91, 2.10, 2.29]$, where $D(\bar{z}^i)$ is the linear growth factor evaluated at the mean redshift $\bar{z}^i$ for each bin $i$ at our fiducial cosmology. Magnification biases $b_m^i = [-0.898, -0.659, -0.403, -0.0704, 0.416]$ are taken from Ref.~\cite{Fang:2020}, which assumes the sample has been selected based on an $r$-band limiting magnitude of 24.81. 

We generate the redshift distribution shapes in a multi-stage process. First, we produce a smooth distribution for the total galaxy sample with the shape 
\begin{equation}
    n(z) \propto  z^2 {\rm exp}[-(z/z_0)^\alpha],
\end{equation}
using $z_0=0.26$ and $\alpha=0.94$. Then we divide that distribution based on the nominal bin edges. To account for \photoz uncertainties, we convolve this with a Gaussian filter of width $0.03(1+z)$ before normalizing the resulting distributions so that they integrate to one.  We use the {\sc CosmoSIS} \code{smail} module~\cite{cosmosis} to perform these calculations.  Finally, we use this smooth set of distributions as a basis for generating an ensemble of noisy $n(z)$ realizations, as we discuss in more detail in Sec.~\ref{sec:photoz}. We take the average of those realizations to be our fiducial LSST Y1 redshift distributions, as shown in \fig{fig:nzs}. Note for this figure that the DES fiducial is based on actual mean $n(z)$ distributions, whereas LSST is a forecasted mean distribution causing it to be smoother. We confirm that the DESC SRD priors on the mean shift \photoz nuisance parameter are reasonably consistent with the scatter in the mean redshifts of our noise realizations. Following the DESC SRD, we do not vary the stretch parameter.

\section{Forecasting Methods}\label{sec:forecastmethods}

We study the impact of analysis design choices on the precision and accuracy of cosmological parameter constraints using Fisher forecasts (see e.g. Refs.~\cite{StatMethods,MichiganLecNote} for an introduction). For a set of observables $\vec{d}$ the Fisher matrix encodes the curvature of the likelihood $\mathcal{L}\equiv P(\vec{d}|\vec{p})$ over the space of model parameters $\vec{p}$ about the maximum $\vec{p}_0$:
\begin{equation}\label{eq:Fij}
    F_{ij} = -\left.\left\langle\frac{\partial^2\loglike}{\partial p_i\partial p_j}\right\rangle\right|_{\param=\param_0}.
\end{equation}
where the ensemble average is over realizations of the data.
Under the assumption of a Gaussian likelihood, the forecasted parameter constraints are
\begin{align}
     \mathrm{marginalized}&: \sigma (p_i) = \sqrt{F_{ii}^{-1}} \\
     \mathrm{unmarginalized}&: \sigma (p_i) = 1/\sqrt{F_{ii}}.\label{eq:unmargerr}
\end{align}
Gaussian priors on  parameters are incorporated by adding a prior matrix ($F^P_{ij}$) to the Fisher matrix such that
\begin{equation}
    F_{ij} \rightarrow F_{ij} + F^P_{ij}
\end{equation}
where
\begin{equation}
    F^P_{ij} = \frac{\delta_{ij}}{(\sigma_{i}^P)^2}
\end{equation}
with $\sigma_{i}^P$ defined as the width of the prior for the parameter $p_i$.

We define our observables to be the angular power spectra $\mathbf{d}=\hat{C_\ell}^{AB}$, where $(A,B)$ label redshift bins the multipole range described above in Section~\ref{sec:modeling}. We assume the likelihood is a multivariate Gaussian in these observables, with a covariance defined according to \eq{eq:cov} that can be treated as independent of the model parameters. Given this, and defining the model prediction for these observables to be $\vec{m}(\vec{p})$, the Fisher matrix is given by
\begin{equation}\label{eq:fisherdef}
F_{ij} = \sum_{a=1}^{N_d} \sum_{b=1}^{N_d} \frac{\partial m_a}{\partial p_i} \Sigma_{ab}^{-1} \frac{\partial m_b}{\partial p_j}.
\end{equation}
Here the sum over indices $a$ and $b$ iterate over all $N_d$ combinations of redshift bin pairs and multipoles included in the analysis, and $\Sigma_{ab}$ corresponds to the relevant components of $\rm{Cov}(\hat{C}^{AB}_{\ell},\hat{C}^{DE}_{\ell'})$. 

We can also use the Fisher formalism to estimate biases in parameter estimates that can arise from inaccuracies in our model predictions, or equivalently, unmodeled contamination affecting measurements. Defining $\vec{m}^{\rm sys}$ to be the unmodeled residual systematics, the predicted Fisher parameter bias is 
\begin{equation}\label{eq:fisherbiasdef}
     p_i^\mathrm{sys} = \sum_j (F^{-1})_{ij}\, \sum_{a=1}^{N_d} \sum_{b=1}^{N_d} m_a^\mathrm{sys}  \Sigma_{ab}^{-1} \frac{\partial m_b}{\partial p_j}.
\end{equation}
This estimate again relies on the approximation of the likelihood as Gaussian in model parameters $\vec{p}$, and additionally approximates the observables as linear in $\vec{m}^{\rm sys}$. 

\subsection{Aside on Fisher Forecasting Methodology}

We note that in this paper, as in most galaxy survey analyses referenced herein, we perform our forecasts in the so-called estimator perspective, where our observables are defined to be the power spectra  $\dat=\{C^{AB}_{\ell}\}$ for all tracer combinations $A$ and $B$. Because angular power spectra contain complete statistical information for Gaussian fields, and because galaxy number density at large scales is well approximated by a Gaussian field, we could have alternatively performed forecasts in a field perspective that defines our observables $\vec{d}$ to be the spherical harmonic components $\dat=\{a^A_{\ell m}\}$ for all tracers $A$, with the power spectra entering only through the covariance. When considering all auto- and cross-correlations, the field and estimator perspective forecasts will produce identical results. 

This equivalence is not guaranteed to hold when we consider analyses including only auto-correlations. In the estimator perspective this is simple: we remove cross-correlations  $C^{AB}_{\ell}$ with $A\neq B$ from our data vector (equivalently, we marginalize over these measurements), but still account for cross-bin correlations when computing the covariance as in \eq{eq:cov}. Doing this in the field perspective is not straightforward, and we find that our approach to doing this breaks the equivalence between calculations done in the field and estimator perspectives. For instance, we find that switching from the estimator to field perspective would alter some of our forecasts results by about 10\%. 

Again, for all results reported in this paper we work in the estimator picture in order to match the likelihoods used in practice for survey analyses. However, as a matter of interest, in \app{sec:appendix}, we include a general derivation of the Fisher matrix and Fisher bias, as well as a more detailed discussion of the relationship between field and estimator perspectives. 

\subsection{Parameters and Priors}\label{sec:paramsandpriors}
 
Our fiducial cosmology values are listed in \tab{table:fidcosmoparams}. In our forecasts we vary  matter density $\Omega_{\rm c}$ and the amplitude of large-scale matter density fluctuations parameterized by $\sigma_8$, fixing the Hubble parameter $h\equiv H_0/(100\,\text{km}\text{s}^{-1}\text{Mpc}^{-1})$,  baryon density $\Omega_{\rm b}$,  neutrino mass $\sum m_{\nu}$,  curvature density $\Omega_k$, and the slope $n_s$ of the primordial power spectrum to  fiducial values. This is representative of including strong priors from additional cosmological probes, such as CMB anisotropies~\footnote{For DES, varying \{$\Omega_{\rm c}$, $\sigma_8$, $H_0$, $\Omega_{\rm b}$, $n_s$\} with \textit{Planck} priors on \{$H_0$, $\Omega_{\rm b}$, $n_s$\} yields a $\mathcal{O}(10\%)$ difference in the marginalized errors of $\Omega_{\rm c}$ and $\sigma_8$ compared to constraints obtained when these parameters are held fixed. We find that Fisher biases scale similarly, such that the size of the parameter biases relative to uncertainties are comparable.}. 

 \tab{table:fidparams} contains survey-specific parameter values and their corresponding priors. We fix magnification bias following Ref.~\cite{DES:magnification} which showed for DES Y3 that cosmology parameters had no significant change if $b_m$ was varied with a Gaussian prior. Along with the cosmology parameters, this means the full set of parameters varied in our forecasts is \{$\Omega_{\rm c}$, $\sigma_8$, $b_g^i$, $\Delta z^i$, $\sigma_{z}^i$\}. The computation of derivatives with respect to these parameters, which we use in the Fisher matrix/bias equations, is detailed in \app{app:numderiv}. 

As described above in \sects{sec:des} and~\ref{sec:lsst}, the DES priors are chosen to follow the DES Y3 Maglim sample redshift calibration analysis of Ref.~\cite{DES:2022redshiftcal}, while the LSST Y1 choices are based on the DESC SRD~\cite{LSST:DESC_SRD} as well as Refs.~\cite{Fang:2020,Zhang:2022}. For both surveys, two choices of Gaussian priors were considered for linear galaxy bias $b_g$. A wide prior, with Gaussian width equal to 50\% of the fiducial value, represents analyses with unconstrained bias, while a tighter 10\% prior serves as a proxy for how results might change in a combined analysis with weak lensing. All Gaussian priors are centered at our fiducial parameter values.

\begin{table}
\footnotesize
\centering
\begin{tabular}{l|c|c|c|c|c|c|c|c|c}
    Param. & \,$\epsilon$\, & $h$ & $\Omega_{\rm b}$ & $\Omega_{\rm c}$ & $\sum m_{\nu}$ & $\Omega_k$& $10^9A_s$ & $n_s$ & $\sigma_8$ \\
   \hline 
    Fid. val.& 1 & 0.675 & 0.048 & 0.268 & 0.06 eV & 0.0 &  2 & 0.965 & 0.803\\
\end{tabular}
\captionsetup{justification=raggedright}
\caption{Fiducial values of the cosmological parameters used in our forecasts, as well as the $\epsilon$ parameter characterizing the significance of RSD and magnification signals.
}
\label{table:fidcosmoparams}
\end{table}

\begin{table*}
\centering
\begin{tabular}{l|c|c|l}

     & Parameter & Fiducial Values  &  Gaussian Prior Widths \\
    \hline\hline
    \multirow{5}{*}{DES Y3}
    &\multirow{2}{*}{$b_g^{i}$} & \multirow{2}{*}{[1.5, 1.8, 1.8, 1.9]} &   [0.75, 0.9, 0.9, 0.95] (unconstrained) \\ 
    & & &   [0.15, 0.18, 0.18, 0.19] (constrained) \\
    \cline{2-4}
     & $b_m^{i}$  & [0.43, 0.30, 1.75, 1.94] & Fixed  \\  
    \cline{2-4}
    &$\Delta z^{i}$ & [0, 0 ,0, 0] &  $10^{-3}\times$[16.4, 10.0, 8.5, 8.4] \\
    \cline{2-4}
     & $\sigma^i_{z}$  &  [1, 1, 1, 1] &   $10^{-3}\times$[63.9, 62.4, 31.5, 40.9] \\ 
   \hline \hline
    \ \multirow{5}{*}{LSST Y1}
    &\multirow{2}{*}{$b_g^{i}$} & \multirow{2}{*}{[1.56, 1.73, 1.91, 2.10, 2.29]} &   [0.78, 0.87, 0.96, 1.05, 1.15] (unconstrained) \\ 
   & &  &   [0.16, 0.17, 0.19, 0.21, 0.23] (constrained)\\
    \cline{2-4}
    & $b_m^{i}$ &  [-0.898, -0.659, -0.403, -0.0704, 0.416] &  Fixed \\
    \cline{2-4}
    & $\Delta z^{i}$  &  [0, 0, 0, 0, 0] &    $10^{-3}\times$ [6.58, 7.52, 8.50, 9.48, 10.47] \\ 
    \cline{2-4}
    &$\sigma^i_{z}$ & [1, 1, 1, 1, 1] &  Fixed \\
   \hline \hline
\end{tabular}
\captionsetup{justification=raggedright}
\caption{The fiducial values and priors of the survey-dependent parameters used in our Fisher forecast. For each redshift bin $i$, these parameters are   galaxy bias $b_g^{i}$, magnification bias $b_m^{i}$, the \photoz distributions' mean shift   $\Delta z^i$, and the \photoz distributions'  stretch  $\sigma^i_{z}$. Arrays report parameter values ordered from lowest to highest redshift.  All Gaussian priors are centered on the fiducial values, and the widths reported in the rightmost column are those priors' standard deviations. }
\label{table:fidparams}
\end{table*}

\section{The Impact of RSD and Magnification}\label{sec:rsdmag}
 
\begin{figure}
    \centering
    \includegraphics[width=\linewidth]{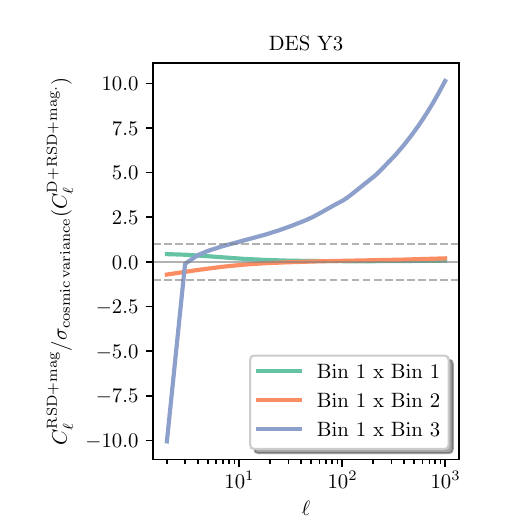}
    \captionsetup{justification=raggedright}
    \caption{DES clustering predictions showing the size of the combined RSD and magnification signal versus the cosmic variance uncertainty for the total signal (density + RSD + magnification). The grey dashed lines encompass the $\pm 1 \sigma$ region, the  auto-correlation signal of bin 1 (the lowest redshift bin) is shown in green, and the orange and purple lines show bin 1's cross-correlations with bins 2 and 3 respectively. With increased redshift separation between galaxies, the combined RSD and magnification contributions become more significant.}
   \label{fig:RSDmagvstot}
\end{figure}

We first characterize the impact of RSD and magnification on our  photometric clustering forecasts, with the goal of evaluating how consideration of these effects influences the decision of whether to include cross-bin correlations in an analysis. As described in \sect{sec:modeling}, RSD and magnification contribute to projected galaxy clustering along with a density term~\footnote{We neglect sub-dominant relativistic contributions to the observed number counts beyond RSD and magnification. For a discussion of the detectability of these effects in future surveys see e.g.~\cite{Alonso_2015}}. While the density term dominates auto-correlations, its contributions decrease for cross-correlations between galaxy samples with limited overlap in co-moving distance. In contrast, RSD and magnification add cosmological signal to cross-correlations: the change in observed redshift due to peculiar velocities can cause a galaxy to be assigned to a different tomographic bin than would have been the case in the absence of RSD, and the magnification signal for different tomographic bins is generated from lensing by the same intervening large-scale structures. By including these effects in our model predictions, we both more accurately account for contributions to observed galaxy clustering and have the potential to extract more information about cosmological structure growth from measurements.

\fig{fig:RSDmagvstot} illustrates the relative significance of these effects for  a selection of DES clustering predictions. For  galaxies in the lowest redshift bin (bin 1), as well as for their cross-correlations with the second and third bins, we show the size of the combined contribution from RSD and magnification relative to  cosmic variance uncertainty for the total (density+RSD+magnification) clustering signal.  We see that this contribution is the smallest for the auto-correlation, and grows as the galaxies are more separated in redshift. For the cross-correlation between the first and third bins, RSD + magnification effects are significantly larger than the cosmic variance uncertainty on the power spectrum.

\subsection{Significance of RSD and Magnification Signals}\label{sec:rsdmag_snr}

 \begin{figure}
    \centering
    \includegraphics[width=\linewidth]{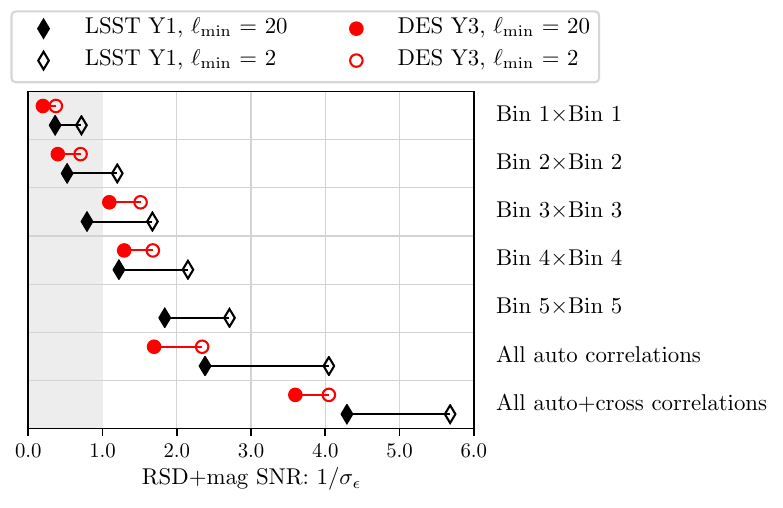}
    \captionsetup{justification=raggedright}
\caption{Values of $1/\sigma_{\epsilon}$, the forecasted signal-to-noise ratio of detecting the presence of RSD and magnification when all model parameters are fixed. Results for DES Y3 and LSST Y1 are shown in red circles and black diamonds, respectively. Filled markers correspond to $\ell_{\rm min}=20$ while open markers show results for  $\ell_{\rm min}=2$. The RSD and magnification signal is detectable if points are outside the gray shaded region. This indicates that the SNR of single-bin-based detections increases with redshift, but are mostly undetectable for both surveys. However, the combined auto-correlation signal is detectable, with the highest SNR obtained when all auto- and cross-correlations are used.}
   \label{fig:epsilonsnr}
\end{figure}

To quantify how detectable RSD and magnification are for a given survey, we perform a Fisher forecast evaluating our sensitivity to a single parameter $\epsilon$ which scales the RSD and magnification contributions to clustering as follows: 
\begin{equation}
    C_\ell^{\mathrm{tot}} = C_\ell^{\mathrm{D}} + \epsilon( C_\ell^{\mathrm{R}} + 2C_\ell^{\mathrm{D}\times\mathrm{R}} + C_\ell^{\mathrm{M}} + 2C_\ell^{\mathrm{D}\times\mathrm{M}} + 2C_\ell^{\mathrm{R}\times\mathrm{M}}).
\end{equation}
Here the $C_\ell$ terms here are defined as in \eq{eq:cellint}, with superscripts referring to contributions from the density (D), RSD (R), and magnification (M) transfer functions of Eqs.~(\ref{eq:DeltaD}--\ref{eq:DeltaM}). If there is only one superscript, the power spectrum component comes from the product of two copies of the same transfer function, whereas the cross ($\times$) terms come from two different transfer functions. We suppress sample (redshift bin) labels for simplicity. 

Our fiducial model corresponds to $\epsilon=1$, and the forecasted error $\sigma(\epsilon) \equiv 1/\sqrt{F_{\epsilon \epsilon}}$ indicates the significance of RSD and magnification contributions to observed clustering. If $\sigma(\epsilon)<1$ we can in principle detect the RSD and magnification effects, and the inverse $1/\sigma(\epsilon)$ can be interpreted as the signal to noise ratio (SNR)  for the presence of RSD and magnification in the observed clustering.  We performed this analysis for auto-correlations for each redshift bin individually,  for the combined constraint from all auto-correlations, and for the combination of all  auto- and cross-correlations.

The resulting SNR values are shown in \fig{fig:epsilonsnr} for DES Y3 and LSST Y1, for two choices of $\ell_{\rm min}$. Focusing first on single bin constraints, we see that the SNR tends to increase with the redshift of the bin, echoing findings from  Ref.~\cite{Euclid:rsd}'s Euclid forecasts. LSST's greater precision, compared to DES, translates into more sensitivity to the RSD and magnification signals, but its single-bin detections remain at low significance. Combining the auto-correlation of multiple redshift bins produces SNR $>1$ for both surveys, and additionally including all cross-bin correlations increases the SNR by a factor of 2.1 for DES and 1.8 for LSST (with $\ell_{\rm min}=20$).  This indicates that about half of the RSD and magnification signal is in the cross-bin correlations.  If we now examine the trend of SNR with $\ell_{\rm min}$, the plot shows that if we can observe lower multipoles, the significance of RSD and magnification increases by up to a factor of two for a given galaxy survey.

\subsection{Impact on Cosmology Inference}\label{sec:impactonconstraints}

\begin{figure*}
     \centering
     \begin{subfigure}[b]{0.49\textwidth}
         \centering
         \includegraphics[width=\linewidth]{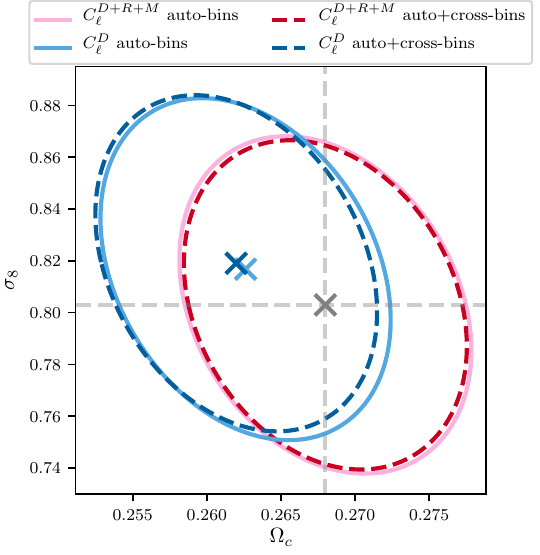}
         \caption{DES}
     \end{subfigure}
     \hfill
     \begin{subfigure}[b]{0.49\textwidth}
         \centering
         \includegraphics[width=\linewidth]{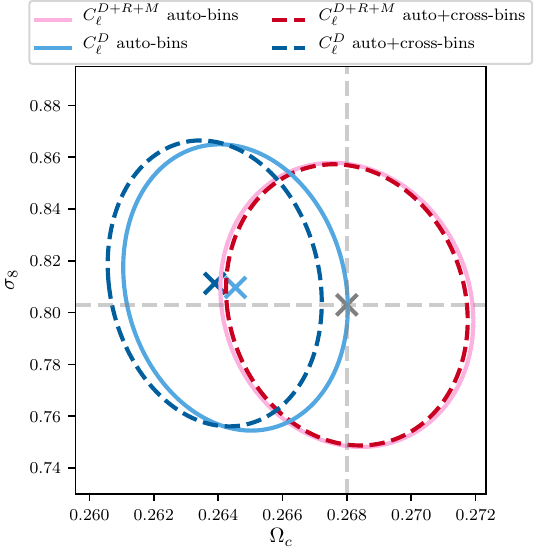}
         \caption{LSST}
     \end{subfigure}
     \captionsetup{justification=raggedright}
     \caption{Forecasted constraints for DES (a) and LSST (b) produced with constrained (10\%) galaxy bias priors. Dashed lines in dark colors correspond to the combination of all auto- and cross-correlations, whereas solid lines in lighter colors are based on auto-correlations only. Blue contours show forecasts in which RSD and magnification are neglected from the model, while pink and red contours correctly account for them. The dashed gray lines show the fiducial (true) parameter values. Including cross-bin correlations essentially has no affect on the cosmology constraints. Neglecting RSD and magnification leads to a significant bias for both DES and LSST.
     }
    \label{fig:DESandLSSTsig8omc}
\end{figure*}

Next, we extend our Fisher forecast to study how incorporating RSD and magnification impacts cosmological constraints. Here, we fix $\epsilon =1$ and vary the parameters $\{\Omega_{\rm c}, \sigma_8, b_g^i, \Delta z^i, \sigma_{z}^i\}$. We forecast constraints with and without including RSD and magnification in our model predictions, and use the Fisher bias expression from \eq{eq:fisherbiasdef} to explore how much neglecting RSD and magnification from our model biases parameter estimates. The systematic effect we are considering is the difference between angular power spectra computed with and without RSD and magnification: 
\begin{equation}
\begin{split}
C_\ell^{\rm sys} &= C_\ell^{\mathrm{tot}} - C_\ell^{\mathrm{D}}\\
&=C_\ell^{\mathrm{R}} + 2C_\ell^{\mathrm{D}\times\mathrm{R}} + C_\ell^{\mathrm{M}} + 2C_\ell^{\mathrm{D}\times\mathrm{M}} + 2C_\ell^{\mathrm{R}\times\mathrm{M}}.
\end{split}
\end{equation}

In \fig{fig:DESandLSSTsig8omc}, we present the results of forecasts for DES (left panel) and LSST (right panel), comparing the constraints based on only auto-correlations (solid lines) and those where all cross-correlations between bins are included (dashed lines). We focus on the results using our constrained galaxy bias prior, which is a 10\% bound on each $b_g^i$ parameter that we use as a proxy for a combined analysis with weak lensing.

We begin by benchmarking our results against published constraints from DES. DES Y3 2$\times$2pt constraints from galaxy clustering combined with galaxy-galaxy lensing~\cite{DES:maglim2x2pt_results} produces parameter errors $\sigma(\Omega_{\rm m})\sim 0.03$ and $\sigma(S_8)\equiv\sigma(\sigma_8(\Omega_{\rm m}/0.3)^{0.5}) \sim 0.04$. This matter density constraint is significantly weaker than our DES auto-only forecasts of $\sigma(\Omega_{\rm m}) =  0.007$, while the $S_8$ bounds are slightly tighter than our $\sigma(S_8) = \pm0.05$. The behavior of the matter density constraints is a consequence of the fact that we fix all cosmological parameters other than $\Omega_{\rm c}$ and $\sigma_8$, while the published DES results marginalize over $\{h, \Omega_{\rm b}, n_s, m_{\nu}\}$ as well as additional nuisance parameters. Our $\Omega_{\rm c}$ bounds are instead more comparable with those from combinations of DES galaxy clustering and weak lensing with external data. For example, analyzing a publicly available chain run for Ref.~\cite{DES:Y3ext}  which fits DES Y3 3$\times$2pt combined with external BAO, RSD, and supernova data with fixed neutrino mass, we find $\sigma(\Omega_{\rm m})={\pm0.007}$\footnote{The data used for this can be found online at \url{https://des.ncsa.illinois.edu/releases/y3a2/Y3key-extensions}. }.  The comparison to our $S_8$ constraints is determined by our choice of galaxy bias priors, since for the analysis of only galaxy clustering, $\sigma_8$ is degenerate with rescaling the galaxy bias parameters. Our galaxy bias priors are weaker than the actual DES Y3 galaxy bias bounds, which translates to our weaker $S_8$ constraint. 

Based on these comparisons, we are satisfied that our Fisher forecasts are indicative of behavior from analyzing galaxy clustering measurements with constraining priors on $h$, $\Omega_{\rm b}$, and $n_s$. As an additional check, we confirm that we obtain similar results when explicitly varying those parameters while assuming priors based on \textit{Planck} CMB constraints. Ultimately, we expect the relative trends in parameter constraints and biases we observe to hold in a full analysis, but we should be mindful that the exact error values are sensitive to choices of fixed survey-dependent and cosmological parameters, $b_g$ priors, and the approximations of the Fisher formalism. 

\begin{figure}
    \centering
    \includegraphics[width=\linewidth]{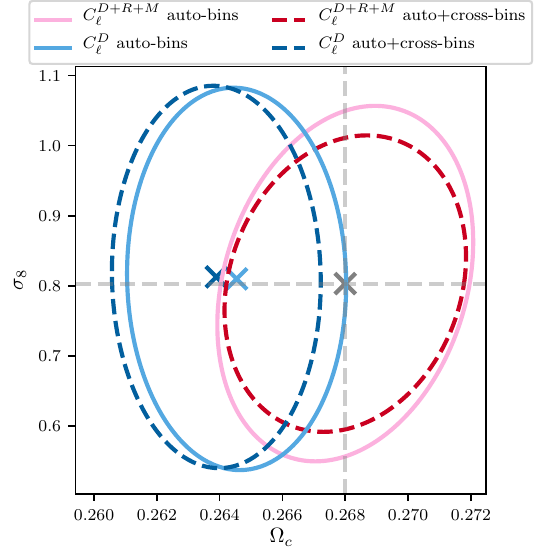}
    \captionsetup{justification=raggedright}
\caption{LSST cosmology constraints with unconstrained (50\%) galaxy bias priors. The auto-correlation constraints are shown in lighter color, solid lines, and the auto- plus cross-correlation constraints have darker colored dashed lines. If RSD and magnification are modelled the contours are in pink and red, if they are neglected they are in blue. Fiducial parameter values are given by the dashed gray lines. For these results with the weaker galaxy bias prior, the combination of RSD, magnification, and cross-bins is able to improve constraints on $\sigma_8$.}
   \label{fig:LSSTbg50sig8omc}
\end{figure}

Now, let us examine the results in \fig{fig:DESandLSSTsig8omc} in more detail, first by comparing auto-only constraints (solid contours) to inferences from the combination of all auto- and cross-correlations (dashed contours). For both surveys, adding cross-bin correlations does not strongly affect bounds on cosmology parameters, when using this constrained galaxy bias prior. However, if we instead use the unconstrained galaxy bias prior, as shown for LSST in \fig{fig:LSSTbg50sig8omc}, then the combination of RSD, magnification, and cross-bin correlations does increase constraining power on $\sigma_8$ (as seen by the dashed red line contour). This implies that the RSD and magnification signal contained in the cross-bins does have additional structure growth information, though this improvement can be overpowered by using tighter galaxy bias priors since they more drastically improve $\sigma_8$ constraints by breaking the $\sigma_8$--$b_g$ degeneracy we mentioned above. Thus this improvement is likely to be subdominant for combined analyses of clustering and weak lensing. 
 
Returning to the constrained galaxy bias prior results in \fig{fig:DESandLSSTsig8omc}, we consider the impact of including RSD and magnification in our model for galaxy clustering (pink and red contours) as opposed to neglecting these effects from our model (blue contours). For DES, neglecting RSD and magnification produces biases in parameter estimates of $\Delta\Omega_{\rm c}=0.8 \sigma \ (1.0 \sigma)$  and $\Delta\sigma_8 = 0.3 \sigma \  (0.4 \sigma)$  for the auto-only (auto+cross) forecasts. While these shifts are $\leq 1\sigma$, we note that they are larger than would likely be considered acceptable for systematic biases from known modeling inaccuracies. For example, DES Y3 modeling validation used a threshold  of $0.3\sigma$ as the target  for acceptable errors from individual systematics~\cite{DES:Y3methods}. For LSST, the $\Omega_{\rm c}$ parameter biases become more severe, with auto-only (auto+cross) shifts of $\Delta\Omega_{\rm c}=1.3 \sigma \ (1.7 \sigma)$ and  $\Delta \sigma_8=0.2 \sigma \ (0.2 \sigma)$. In almost all cases, the biases from neglecting RSD and magnification in our model are larger when cross-bin correlations are included, reflecting the fact that the cross-correlations enhance the significance of the RSD and magnification signals. 

These results inform two key observations. First, they demonstrate that it is best practice when modeling photometric clustering for Euclid, Roman, or LSST to incorporate RSD and magnification because neglecting them will bias parameter estimates, in line with previous studies in the literature~(e.g.~Refs.~\cite{DES:Y3methods,Mahony:magnification,Lorenz:2017iez,TanidisCamera:rsd:2019teo,Tanidis:mag:2019fdh}). Additionally --- and also in line with Refs.~\cite{TanidisCamera:rsd:2019teo,Tanidis:mag:2019fdh} --- when RSD and magnification are modeled, adding cross-bin correlations can break degeneracy between $\sigma_8$ and galaxy bias, but more weakly than what is expected from combined analyses with weak lensing. In other words: while modeling these effects is necessary for unbiased inference, their inclusion in forecasts will not substantially affect the evaluation of whether or not it is worth including cross-bin correlations in an analysis.

\section{The Impact of Photometric Redshift Uncertainties}\label{sec:photoz}

Our model for projected galaxy clustering assumes a fiducial redshift distribution $n_i(z)$ for each galaxy sample $i$. This is an estimate of the sample's true underlying distribution that will necessarily have some level of inaccuracy due to \photoz errors. The goal of marginalizing over the \photoz shift $\Delta z^i$ and stretch $\sigma_z^i$ nuisance parameters is to absorb the effects of discrepancies between the estimated and true $n(z)$ distributions so that those mismatches do not bias our cosmological parameter estimates. Here, we examine whether including cross-bin correlations impacts the efficacy of this treatment, assessing how the potential for enhanced precision compares to the risk of systematic biases from redshift uncertainties. 

To see why this is a concern, we note that in \eqs{eq:cellint}--(\ref{eq:DeltaD}), the dominant density term in the cross-correlation between two galaxy samples depends on an integral over the product of their redshift distributions. For auto-correlations, this integral over $n_i^2(z)$ receives most of its contributions from redshifts near the peak of the sample's redshift distribution, and so is relatively insensitive to the shape of the $n(z)$ tails. This means that most of the impact of \photoz uncertainties can be captured by marginalizing over an overall shift or stretch of the redshift distribution. For cross-correlations, however, the projection integral of \eq{eq:cellint} depends on $n_i(z)n_j(z)$ for $i\neq j$ and thus only receives contributions from where the two distributions overlap. Especially for well-separated redshift bins, the integral becomes sensitive to the shape of the $n(z)$ tails in a way that may not be fully captured by simply shifting and stretching the distributions.

\subsection{Realistic \hmath $n(z)$ Noise Realizations} 

\begin{figure}
    \begin{subfigure}[b]{0.49\textwidth}
         \centering
         \includegraphics[width=\linewidth]{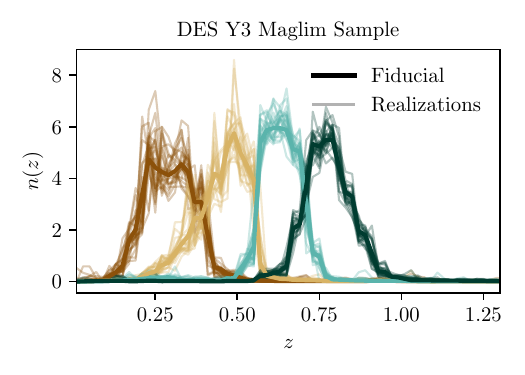}
         \label{fig:DESnzens}
     \end{subfigure}
     \hfill
     \begin{subfigure}[b]{0.49\textwidth}
         \centering
         \includegraphics[width=\linewidth]{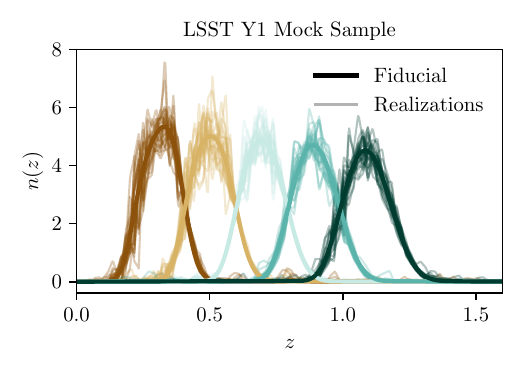}
         \label{fig:LSSTnzens}
     \end{subfigure}
    \captionsetup{justification=raggedright}
    \caption{An ensemble of 25 $n(z)$ realizations (faded lines) compared to the fiducial $n(z)$ (solid line) for DES (\textit{top panel}) and LSST (\textit{bottom panel}). The DES realizations are from \cite{DES:2022redshiftcal} and were adapted to create the LSST realizations using the procedure described in the text. }
    \label{fig:nzensemble}
\end{figure}

In order to assess the extent to which inaccuracies in our modeled redshift distributions due to \photoz errors impact cosmological constraints, we perform Fisher bias forecasts using an ensemble of realistic $n(z)$ realizations. As described above in \sect{sec:des}, we use a subset of an ensemble of 6000 realizations produced as part of the DES \maglim \photoz calibration study reported in Ref.~\cite{DES:2022redshiftcal}. This ensemble reflects our best estimate of the distribution of possible $n(z)$ variations, and is the product of a self-organizing map analysis of the galaxies' photometry calibrated against additional color measurements and cross-correlations with galaxy samples.  For DES forecasts, we randomly select 100 $n(z)$ realizations from this ensemble, a subset of which are shown in the upper panel of \fig{fig:nzensemble}, and use them to compute 100 sets of predicted clustering power spectra $C_\ell^{\rm r}$, where $r$ labels the realization. We then use each $C_\ell^{\rm r}$ to forecast how cosmological parameter estimates are biased if the true redshift distribution corresponds to the realization while our model fit is performed assuming the fiducial distribution (which is the average of all the $n(z)$ realizations). In other words, we perform a Fisher bias forecast using the data contamination: 
\begin{equation}
\begin{split}
C_\ell^{\rm sys} &= C_\ell^{\mathrm{r}} - C_\ell^{\mathrm{fid}}.
\end{split}
\end{equation}

For LSST, we conduct a similar analysis using a set of mock $n(z)$ realizations that we produce based on the DES ensemble as follows. As described above in \sect{sec:lsst}, we begin by creating smooth redshift distributions for the LSST clustering sample $n_s^{\rm LSST}(z)$, following LSST DESC SRD specifications. (Note that though we suppress tomographic bin labels for conciseness in the above discussion, we do perform this procedure for all tomographic bins.) Then, for each DES realization $n_r^{\rm DES}(z)$, we extract the difference from the fiducial distribution,
\begin{equation}\label{eq:noise}
    d^{\rm DES}_r(z)=n_r^{\rm DES}(z)-n_{\rm fid}^{\rm DES}(z).
\end{equation}
Next, to roughly preserve the distribution of $n(z)$ features relative to bin centers and widths, we rescale these noise differences so that
\begin{equation}
    d^{\rm LSST}_r(z) = d^{\rm DES}_r\left(\left[z-\bar{z}_{\rm fid}^{\rm DES}\right]\left[\frac{w^{\rm LSST}_s}{w^{\rm DES}_{\rm fid}}\right] + \bar{z}^{\rm LSST}_s\right).
\end{equation}
Here, $\bar{z}^{\rm DES}_{\rm fid}$ is the mean redshift of the fiducial DES distribution, $w^{\rm DES}_{\rm fid}$ is the width in redshift of the region where $n_{\rm fid}^{\rm DES}(z)>0.1$, while $\bar{z}^{\rm LSST}_s$ and $w^{\rm LSST}_s$ are the same quantities defined for the smooth LSST distribution $n_s^{\rm LSST}(z)$. Finally, we add this scaled noise realization to the smooth distribution, which we normalize such that
\begin{equation}
    n_r^{\rm LSST}(z) = \frac{n_s^{\rm LSST}(z) + d^{\rm LSST}_r(z)}{\int dz (n_s^{\rm LSST}(z) + d^{\rm LSST}_r(z))}
\end{equation}
A subset of these realizations are plotted in the lower panel of \fig{fig:nzensemble}. To ensure the resulting redshift distributions are in line with what we might expect for forecasted LSST \photoz uncertainties, we compare the scatter in the realizations' mean redshifts to the DESC SRD prior on the \photoz mean shift nuisance parameter. We discard any realizations with a mean scatter larger than 5 times the prior width. We then define our fiducial LSST redshift distribution to be equal to the average of the remaining realizations.  Of the redshift distributions selected in this way, we find the mean redshift of each bin to have standard deviations [0.0092, 0.0073, 0.0066, 0.0086, 0.0092], which are comparable to the $\Delta z$ prior widths of [0.0066, 0.0075, 0.0085, 0.0095, 0.0100].

\subsection{Noise Bias Correction}\label{sec:noisebias}
 
To put our Fisher bias estimates from different $n(z)$ realizations into context, we note that slightly altering the definition of our fiducial redshift distribution can more accurately capture how \photoz uncertainties propagate to impact our expected parameter constraints. To see this, it is worthwhile to first build some intuition for the relationship between variations in $n(z)$ realizations and the scatter they produce in angular power spectra predictions. Specifically, we consider the impact of the noise bias discussed in Refs.~\cite{ACTFarren:2023oei,Krolewski:2021yqy}. This effect comes from the fact that we set our fiducial redshift distribution equal to the mean of an ensemble of possible realizations, $n^i_{\rm fid} \equiv \langle n^i_r\rangle$, while angular power spectra have a quadratic dependence on the distributions, $C_{\ell}^{ij}\sim n^i(z)n^j(z)$. Simply put, the fact that $\langle n^2(z) \rangle \neq \langle n(z)\rangle^2$, where angled brackets are averages over the realizations, will cause the angular power spectra computed  using that fiducial distribution to be biased relative to the ensemble of power spectra computed directly from those $n(z)$ realizations.

Here, we  estimate the size of this noise bias  both to characterize how it contributes to our Fisher bias estimates, and to guide whether it is significant enough to motivate updates in how future surveys define their fiducial $n(z)$ given \photoz uncertainties. To do this, we compare our fiducial angular power spectra calculations (using $\langle n(z)\rangle^2$) to a version that more accurately captures how the scatter in \photoz realizations translates to an average power spectrum prediction by instead using $\langle n(z)^2 \rangle$ for each tomographic bin. More specifically, we input the function, $\sqrt{\langle n(z)^2 \rangle}$, unnormalized, as redshift distributions when computing angular power spectra with \code{CAMB}. We highlight that this alternative calculation could be a promising method for future surveys to reduce the impact of noise bias with very little additional computational overhead. 

\begin{figure}
    \centering
    \includegraphics[width=\linewidth]{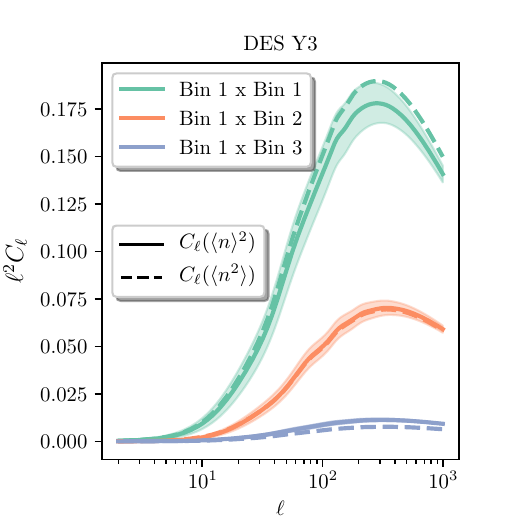}
    \captionsetup{justification=raggedright}
    \caption{The difference between DES $C_\ell$ calculations using our fiducial calculation, based on  the average of $n(z)$ realizations ($C_\ell(\langle n(z) \rangle^2)$), and an alternative calculation correcting for $n(z)$ noise bias ($C_\ell(\langle n(z)^2 \rangle)$). The shaded regions are the cosmic variance error bars for $C_\ell(\langle n(z) \rangle^2)$, note that the purple error bars are too small to be seen on this plot. This shows that using $C_\ell(\langle n(z) \rangle^2)$ causes a low bias of the auto-correlation power spectrum in comparison to $C_\ell(\langle n(z)^2 \rangle)$.
    }
   \label{fig:n2avgvsavgfirst}
\end{figure}

\fig{fig:n2avgvsavgfirst} compares our fiducial definition of $C_\ell(\langle n(z) \rangle^2)$, plotted with solid lines, to the noise-bias-corrected alternative $C_\ell(\langle n(z)^2 \rangle)$, plotted with dashed lines. We see that the fiducial prediction for the auto-correlation power spectrum is lower than the alternative method by about $1\sigma$ compared to cosmic variance error bars.  For cross-correlation  between neighbouring bins, such as the orange 1$\times$2 lines shown in the plot, we see relatively little effect. This means $\langle n^i n^j \rangle\approx\langle n^i\rangle\langle n^j \rangle $, suggesting that the noise in the redshift distributions is uncorrelated between the tomographic bins. We see a more significant change relative to cosmic variance for the well-separated 1$\times$3 bin pair in the plot, but note that the relatively small amount of power in that cross-correlation compared to auto-spectra and nearest-neighbor pairs means it is less likely to impact cosmological inference. We see similar behavior for LSST.
 
We perform our forecasts using the standard $\langle n(z) \rangle^2$ calculation to match what is done in practice in survey analyses, but we argue that the alternative $\langle n(z)^2 \rangle$ treatment is the preferable approach for propagating $n(z)$ uncertainties to power spectra expectation values. We characterize how much this choice impacts cosmological parameter estimates via a Fisher bias estimate assuming:
\begin{equation}\label{eq:sys_noisebias}
\begin{split}
C_\ell^{\rm sys} &= C_\ell(\langle n(z)^2 \rangle) - C_\ell(\langle n(z) \rangle^2).
\end{split}
\end{equation}
This will tell us how much neglecting noise bias impacts cosmological constraints, as well as the extent to which correcting for noise bias might alleviate  parameter shifts generated by our ensemble of noisy $n(z)$ realizations.

\subsection{Results}

\begin{figure*}
     \centering
     \begin{subfigure}[b]{0.49\textwidth}
         \centering
         \includegraphics[width=\linewidth]{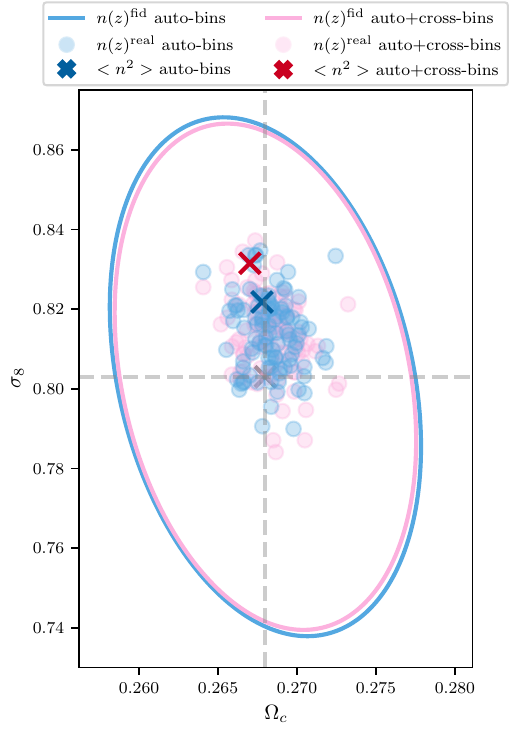}
         \caption{DES}
     \end{subfigure}
     \hfill
     \begin{subfigure}[b]{0.49\textwidth}
         \centering
         \includegraphics[width=\linewidth]{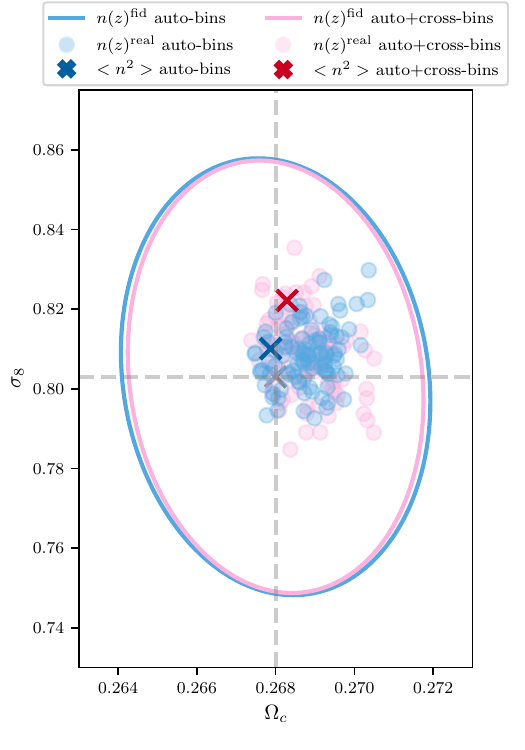}
         \caption{LSST}
     \end{subfigure}
     \captionsetup{justification=raggedright}
     \caption{These are the cosmological constraints for DES (a) and LSST (b) for a constrained galaxy bias prior. In blue are the results when considering only auto-correlations between tomographic bins, and in pink are auto- plus cross-correlations. The solid contours show the constraints from our fiducial $n(z)$ which appears in the $C_\ell$'s as $\langle n \rangle^2$. Then the scatter of circular translucent points illustrates the bias in the cosmology if a $n(z)$ realization is used instead of the fiducial, and the dark crosses are the bias from using $C_\ell(\langle n^2 \rangle)$. When cross-bins are included in the analysis, the parameter biases are only slightly larger than auto-only. Correcting for the noise bias may account for some of the \photoz biases, but will not fully alleviate them.}
    \label{fig:DESandLSSTbiasscatter}
\end{figure*}

\fig{fig:DESandLSSTbiasscatter} shows, for a constrained galaxy bias prior, the cosmological parameter biases induced by mismatches between our fiducial redshift distribution and 100 different ``true'' $n(z)$ realizations. Results are shown for the analysis of only auto-correlations in blue, and for the combination of all auto- and cross-correlations in pink. Circular data points show Fisher biases computed for each of the redshift realizations, ``X'' markers show the effect of noise bias from \eq{eq:sys_noisebias}, and contours are forecasted $1\sigma$ parameter errors. 

We see that parameter biases are only slightly larger for auto+cross constraints than for the auto-only analyses, both in absolute parameter shifts and relative to their parameter constraints. For DES, the scatter of parameter biases are all contained within the $1\sigma$ contours for both scenarios, but we note that they are larger than would be ideal for a more stringent  accuracy requirement:  58\% of the auto-only,  and 62\% of auto+cross Fisher biases are larger than $0.3\sigma$. For LSST, the results are similar to DES, with all biases from $n(z)$ variations within $1\sigma$, and for the auto-only (auto+cross) 61\% (68\%) are larger than $0.3\sigma$.

While not shown in \fig{fig:DESandLSSTbiasscatter}, we also study the constraints and biases on the \photoz nuisance parameters. For auto-only, the \photoz parameters' contours are fully set by the priors, but adding cross-correlations to the analysis significantly tightens constraints on the $\Delta z^i$ shift parameters for DES and LSST, while producing a smaller improvement on the $\sigma_z^i$ stretch constraints (which are only modeled for DES). 
This behavior is consistent with the findings in Refs.~\cite{Nicola:HSCclustering,Schaan:2020qox} that including cross-bin correlations improves constraints on \photoz nuisance parameters. Looking at  Fisher biases, for both DES and LSST auto-only analyses we find that \photoz nuisance parameter biases are smaller than the expected errors. Adding cross-correlations produces changes of the order of $1\sigma$ in the $\Delta z^i$ shift parameters, with larger biases for LSST than DES, but produces little change in the $\sigma_z^i$ values varied in the DES analysis.

These results, along with the fact that the shifts in $\Omega_c$ and $\sigma_8$ are relatively small, suggest that the shift ($\Delta z^i$) nuisance parameters are absorbing most of the impact of $n(z)$ variations and are protecting the cosmology parameters from associated biases. 
We note that though the $\Delta z^i$ Fisher biases are about $1\sigma$ when cross-bins are included, this is not a sign of a flawed analysis. The  $n(z)$ realizations may actually have different means than our fiducial distribution, and so effectively have true shift parameter values that differ from the fiducial $\Delta z^i=0$. When cross-bins are analyzed, we have enhanced sensitivity to bins' mean redshifts, which means we not only tighten $\Delta z^i$ constraints, but also better detect changes in a realization's mean redshift. The fact that this does not produce larger changes in cosmology parameter estimates  indicates that the \photoz nuisance parameter prescription used in our forecasts is equally successful at protecting results from \photoz systematics for the auto-only and auto+cross analyses. This has encouraging implications for future survey analyses, in that it suggests that tools developed to accommodate \photoz uncertainties for auto-correlation-only analyses are likely to remain applicable when cross-correlations are included. Of course, depending on thresholds for acceptable parameter biases and the specifics of a given analysis, future studies should still carefully assess these biases and ensure that their modeling of \photoz uncertainties is sufficiently flexible to accommodate realistic variations in galaxy redshift distributions.

The noise bias results, shown with X markers in \fig{fig:DESandLSSTbiasscatter}, suggest that using the noise-bias-corrected ($\sqrt{\langle n(z)^2 \rangle}$-based) model power spectra can help alleviate parameter biases from $n(z)$ uncertainties, but will not entirely remove the risk. The resulting parameter shifts for DES are $(\Delta\Omega_{\rm c}, \Delta\sigma_8) = (0.03\sigma, 0.44\sigma )$ for the auto-only analysis and $(0.16\sigma,0.68 \sigma )$ for auto+cross, while for LSST the shifts are for $(0.05\sigma, 0.20 \sigma)$ for auto-only and $(0.12\sigma, 0.54 \sigma)$ for auto+cross. 

Recall that these markers show parameter shifts caused by the fact that our fiducial $n_{\rm fid}(z) = \langle n(z)\rangle $  calculations are offset relative to how the ensemble of possible $n(z)$ realizations actually affect the power spectrum expectation values. Roughly speaking, we thus expect the noise bias X's to correspond to the average behavior of the set of realization-specific Fisher biases. For the most part this behavior is reflected in our findings, though the noise biases are somewhat offset from the center of the cloud of points. This is possibly a consequence of the fact that  we are looking at marginalized parameter constraints, so part of the $C_{\ell}$ offsets are being absorbed by parameters other than $\Omega_c$ and $\sigma_8$, as well as sample variance between the 100 plotted realizations and the full ensemble of 6000 used to compute noise bias corrections. Nonetheless, we see that the size of the noise bias shifts are comparable to the scatter from individual $n(z)$ realizations, so can correct for part of the risk of systematic biases from \photoz uncertainties, but not all of it. 

Overall, the primary conclusion of this investigation is that sensitivity to $n(z)$ shape variations is an important consideration when assessing an analysis's systematic error budget. 
To avoid introducing additional bias to the analysis, the fiducial $n(z)$ should be defined such that it accurately propagates the impact of redshift uncertainties on the power spectra expectation values.  
We also find that the level of parameter biases induced by \photoz uncertainty are similar for DES and LSST, as well as for auto-only and auto+cross,  suggesting that variations in the shape $n(z)$ tails are not a limiting systematic for the inclusion of cross-bins in these analyses. Considering the auto-only and auto+cross analyses together, whether the parameter biases we see here are concerning, and thus whether the standard shift-stretch nuisance parameter method we employ will be sufficient for protecting future analyses against \photoz systematics --- as opposed to a more sophisticated  method like those in Refs.~\cite{DES:hyperrank,Schaan:2020qox,Hadzhiyska:2020xob} --- is something that should be tested on a case-by-case basis. This assessment will depend on how  priors and other choices used in our forecasts compare to the specifics of a given analysis, what is considered an acceptable threshold for systematic-driven parameter biases, as well as other developments in \photoz methodology.
In any case, it is clear that testing for robustness to \photoz distribution uncertainties will be an important step of future  photometric galaxy clustering studies. 

\section{Conclusions}\label{sec:conclusion}

In this paper, we examine the benefits and risk of including cross-correlations between different tomographic bins in the cosmological analysis of photometric galaxy clustering. We do this using Fisher forecasts for surveys with properties corresponding to DES Y3 and LSST Y1, in an idealized scenario where we fix cosmological parameters other than $\Omega_{\rm c}$ and $\sigma_8$, but retain variations in galaxy bias and nuisance parameters related to \photoz uncertainties. We focus in particular on how the decision to include cross-bin correlations in an analysis may be influenced by the incorporation of RSD and magnification into our model predictions, as well as by uncertainties in the shape of the galaxy redshift distributions. 

Modeling RSD and magnification contributions to photometric clustering have only recently been required due to improvements in galaxy survey precision, and they enhance cross-correlations between redshift bins in a way that carries information about the growth of large scale structure. We demonstrate that cross-bin correlations contain about half of the SNR for the detection of RSD and magnification in photometric clustering, and, in line with previous results in the literature, that neglecting these effects from model predictions biases cosmological inferences. If a constraining galaxy bias prior is used, then we do not see a difference in cosmological constraining power from adding in cross-correlations whether or not RSD and magnification is modeled. However, if the galaxy bias prior is unconstrained, then we can see an improvement in $\sigma_8$ constraints from including the combination of RSD, magnification, and cross-correlations which break the degeneracy between $\sigma_8$ and galaxy bias. Therefore, there is additional structure growth information to be obtained from the cross-bin RSD and magnification signals, but it is sub-dominant to the effects of improvements to galaxy bias constraints, eg. via a combined analysis with weak lensing. Given this, results from forecasts done to guide analysis design decisions about cross-bin correlations remain informative even if they did not account for RSD and magnification, but these effects should ultimately be modeled in the analysis of real data. 

To characterize the impact of systematic uncertainties in galaxy redshift distributions, we assess parameter biases caused by differences between our model's fiducial redshift distribution and an ensemble of realistic $n(z)$ realizations obtained from a DES \photoz calibration study. We also discuss the fact that defining our fiducial redshift distribution to be the average of the noisy $n(z)$ realizations will underpredict the expectation value of galaxy auto-power spectra, and we propose a computationally efficient method to correct for this noise bias. For DES and LSST forecasts, we find that the ensemble of $n(z)$ realizations produces biases in $\Omega_{\rm c}$ and $\sigma_8$ that are smaller than the parameters' $1\sigma$ errors, but may be significant enough to be a concern for more stringent accuracy requirements. The inclusion of cross-bins in principal increases sensitivity to the shape of $n(z)$ tails, but in practice it only slightly increases cosmological parameter biases. Therefore, for our forecast set-up and for LSST Y1's level of precision, standard nuisance-parameter methods for protecting cosmological parameter estimates from \photoz uncertainties adequately protect against $n(z)$ biases for both auto-only or auto+cross analyses. Additionally, cross-correlations can self-calibrate \photoz systematics by providing information about the relative bin positions which tightens nuisance parameter constraints (as encountered before by \cite{Nicola:HSCclustering, Schaan:2020qox}). 
Since concern about sensitivity to \photoz uncertainties was a primary motivation for leaving cross-correlations out of analyses, the robustness we find here is encouraging for future surveys.

Ultimately, we conclude that it is worthwhile to include cross-correlations in future analyses because they can help self-calibrate associated systematic uncertainties without significantly increasing the risk of systematic biases. Our findings motivate several promising directions for future investigation.
First, we demonstrate that adding cross-correlations to our analysis does not increase the bias in cosmological parameters from neglecting RSD and magnification, but it does enhance our sensitivity to the presence of these signals. 
It would be interesting to study whether cross-correlations' sensitivity to RSD and magnification become more important for breaking degeneracies in models where physics beyond \lcdm modifies structure growth, such as those studied in e.g.~Refs.~\cite{DES:Y3ext,Ishak:2019aay,Garcia-Garcia:2021unp}.
We also highlight that care must be taken in defining fiducial galaxy redshift distributions so that they accurately capture how \photoz uncertainties affect our expectations for galaxy power spectra, and in ensuring \photoz nuisance parameters can adequately capture possible variations in the underlying $n(z)$ distributions. As data precision and \photoz methods evolve, it will be worth revisiting these considerations to assess whether standard nuisance parameter approaches can effectively separate cosmological and \photoz information, or whether more sophisticated approaches like those of Refs.~\cite{DES:hyperrank,Schaan:2020qox, Hadzhiyska:2020xob} are needed. Finally, while we have shown that for near-future precision  auto+cross and auto-only are comparably susceptible to bias from \photoz systematics, this may not be the case for other astrophysical or measurement uncertainties. For example, it will be important to test the sensitivity of results to assumptions about magnification bias,  building on Ref.~\cite{DES:magnification}, as well as to redshift variations of magnification and galaxy bias parameters as studied in Ref.~\cite{Pandey:2023tjn}.  Pursuing these investigations will support the overarching goal of fully and accurately leveraging the information  galaxy surveys can provide about the Universe.

\begin{acknowledgments}

The authors would like to thank Giulia Giannini for providing data and advice on working with $n(z)$ realizations for the DES Y3 Maglim sample and Judit Prat for providing chains associated with Ref.~\cite{LSSTPratZuntz:2022sql} for LSST forecasting cross-checks. We also thank Noah Weaverdyck, Gerrit Farren, Joe Zuntz,  Agn\`es Fert\'e, Jack Elvin-Poole, and Paul Rogozenski for helpful discussions, as well as Javiera Hern\'adez Morales for participating in the early stages of this project.
Research at Perimeter Institute is supported in part by the Government of Canada through the Department of Innovation, Science and Economic Development and by the Province of Ontario through the Ministry of Colleges and Universities.
JK acknowledges support from the Natural Sciences and Engineering Research Council
of Canada (NSERC) through the Vanier Canada Graduate Scholarship. JM is supported by the BMO Inclusive Excellence Post-Doctoral Fellowship. MCJ is supported by the Natural Sciences and Engineering Research Council through a Discovery grant. This project made use of the software tools  {\sc CAMB}~\cite{CAMB,Lewis:1999bs,Howlett:2012mh,Challinor:2011bk}, {\sc Numpy}~\cite{numpy}, {\sc Matplotlib}~\cite{matplotlib}, and {\sc SciPy}~\cite{Scipy}. Some of the calculations were also done using the Symmetry computing cluster at Perimeter Institute.  
\end{acknowledgments}

\appendix
\begin{widetext}
\section{Fisher Formalism} \label{sec:appendix}
In this appendix we present, for pedagogical purposes, a derivation of the Fisher matrix and bias from both field and estimator perspectives. We begin by deriving the general Fisher equations in Sec.\ref{sec:derivefisher}, then we find the Fisher matrix and bias for the field (Sec.\ref{sec:field}) and estimator (Sec.\ref{sec:est}) perspectives. Finally, we discuss in Sec.\ref{sec:fieldest} how the two perspectives differ and how that may affect results when performing an analysis using only auto-bin correlations.

\subsection{Derivation of Fisher Matrix and Bias} \label{sec:derivefisher}
First, let us assume the likelihood \like is Gaussian in $\mathbf{d}=\{d_i\}$, which is reasonable for most cosmological observations under the central limit theorem. This gives the likelihood the form
\begin{equation}\label{eq:loglike}
\loglike(\dat|\param) = -\tfrac{1}{2}[\dat-\model]^T\invcov[\dat-\model] -\tfrac{1}{2}\ln{(2\pi)} -\tfrac{1}{2}\ln{[\det{\cov}]}
\end{equation}
where $\mathbf{m}=\langle\mathbf{d}\rangle$ are the theoretical observable quantities (the model) which are calculated at the assumed parameter values $\mathbf{p}$, and $\mathbf{\Sigma}$ is the covariance matrix. 

The key assumption of Fisher forecasting is that the likelihood is {\em also} Gaussian in the parameters \param, with its maximum at $\param_0$. This means we can Taylor expand
\begin{equation}
    \loglike(\dat|\param) \approx \loglike(\dat|\param_0) + \tfrac{1}{2} \sum_{i=1}^{N_p}\sum_{j=1}^{N_p}[\param-\param_0]_i \left. \frac{\partial^2\loglike}{\partial p_i\partial p_j}\right|_{\param=\param_0}
    [\param-\param_0]_j, \label{eq:fishertaylor}
\end{equation}
where we have used the fact that since $\param_0$ maximizes \loglike, the derivative $\partial \loglike / \partial p_i = 0$. Then, the Fisher matrix $F_{ij}$ is defined as in Eq(\ref{eq:Fij}).

To derive the Fisher matrix, it will be easier if we write out matrix multiplication as sums over indices
\begin{equation}
\loglike = -\tfrac{1}{2}\sum_{a=1}^{N_d}\sum_{b=1}^{N_d}[d_a-m_a]\Sigma^{-1}_{ab}[d_b-m_b] -\tfrac{1}{2}\ln{(2\pi)} -\tfrac{1}{2}\ln{\det{\cov}}. 
\end{equation}
Taking one derivative,
\begin{equation}
\begin{split}
\frac{\partial\loglike}{\partial p_i} 
&= -\tfrac{1}{2}\sum_{a=1}^{N_d}\sum_{b=1}^{N_d}\left\{-\frac{\partial m_a}{\partial p_i}\Sigma^{-1}_{ab}[d_b-m_b] - [d_a-m_a]\Sigma^{-1}_{ab}\frac{\partial m_b}{\partial p_i}\right\}\\ 
&\;\;\;\; -\tfrac{1}{2}\sum_{a=1}^{N_d}\sum_{b=1}^{N_d}[d_a-m_a]\frac{\partial \Sigma^{-1}_{ab}}{\partial p_i}[d_b-m_b]-\tfrac{1}{2}\frac{[\det{\cov}]}{[\det{\cov}]} \Tr{\left[\invcov\frac{\partial\cov}{\partial p_i}\right]}\\
    &= \sum_{a=1}^{N_d}\sum_{b=1}^{N_d}\left\{\frac{\partial m_a}{\partial p_i}\Sigma^{-1}_{ab}[d_b-m_b] 
     +  \tfrac{1}{2}[d_a-m_a]\Sigma^{-1}_{ac}\frac{\partial \Sigma_{cd}}{\partial p_i}\Sigma^{-1}_{db}[d_b-m_b]-\tfrac{1}{2} \Sigma^{-1}_{ab}\frac{\partial\Sigma_{ba}}{\partial p_i}\right\} .
\end{split}
\label{eq:singleder}
\end{equation}
To get this we used a couple of matrix identities:
\begin{align}
    \frac{d \invcov}{d \theta} &= - \invcov \frac{d \cov}{d \theta} \invcov\\
    \frac{d [\det{\cov}]}{d\theta} &= [\det{\cov}] \Tr{\left[\invcov\frac{d\cov}{d\theta}\right]}.
\end{align}
Now go back to Eq.~\ref{eq:singleder} and  take another derivative:
\begin{equation}
\begin{split}
\frac{\partial^2\loglike}{\partial p_i\partial p_j} 
&= \sum_{a=1}^{N_d}\sum_{b=1}^{N_d}\left\{ \left(\frac{\partial^2 m_a}{\partial p_i\partial p_j}\Sigma^{-1}_{ab} - \frac{\partial m_a}{\partial p_i}\Sigma^{-1}_{ac}\frac{\partial \Sigma_{cd}}{\partial p_j}\Sigma^{-1}_{db}\right)[d_b-m_b] -\frac{\partial m_a}{\partial p_i}\Sigma^{-1}_{ab}\frac{\partial m_b}{\partial p_j} -  \tfrac{1}{2}\frac{\partial m_a}{\partial p_j}\Sigma^{-1}_{ac}\frac{\partial \Sigma_{cd}}{\partial p_i}\Sigma^{-1}_{db}[d_b-m_b] \right. \\
&\;\;\;\; - \tfrac{1}{2}[d_a-m_a]\left( \Sigma^{-1}_{ac}\frac{\partial \Sigma_{cd}}{\partial p_j}\Sigma^{-1}_{de}\frac{\partial \Sigma_{ef}}{\partial p_i}\Sigma^{-1}_{fb} - \Sigma^{-1}_{ac}\frac{\partial^2 \Sigma_{cd}}{\partial p_i\partial p_j}\Sigma^{-1}_{db} + \Sigma^{-1}_{ac}\frac{\partial \Sigma_{cd}}{\partial p_i}\Sigma^{-1}_{de}\frac{\partial \Sigma_{ef}}{\partial p_j}\Sigma^{-1}_{fb} \right)[d_b-m_b]\\
&\;\;\;\; - \left. \tfrac{1}{2}[d_a-m_a]\Sigma^{-1}_{ac}\frac{\partial \Sigma_{cd}}{\partial p_i}\Sigma^{-1}_{db}\frac{\partial m_a}{\partial p_j} + \tfrac{1}{2}\Sigma^{-1}_{ac}\frac{\partial \Sigma_{cd}}{\partial p_j}\Sigma^{-1}_{de}\frac{\partial \Sigma_{eb}}{\partial p_i} - \tfrac{1}{2}\Sigma^{-1}_{ac}\frac{\partial^2 \Sigma_{cb}}{\partial p_i\partial p_j} \right\}.
\end{split}
\label{eq:doubleder}
\end{equation}

Next, let us take the expectation value and evaluate at $\param_0$.  Here, we will use the definition of the covariance, $\langle(d_a-m_a)(d_b-m_b)\rangle = \Sigma_{ab}$, to turn these terms into traces over covariances. Also note that since $\langle\dat\rangle = \model(\param_0)$, this will cause terms with $\langle \dat - \model \rangle$ to cancel out leaving us with
\begin{equation}
\begin{split}
2\left.\left\langle\frac{\partial^2\loglike}{\partial p_i\partial p_j}\right\rangle\right|_{\param=\param_0} & 
= - 2\Tr\left[ \frac{\partial \model^T}{\partial p_i}\invcov \frac{\partial \model}{\partial p_j}\right] -\Tr{\left[\invcov\frac{\partial \cov}{\partial p_j}\invcov\frac{\partial \cov}{\partial p_i}\invcov\cov\right]} + \Tr{\left[\invcov\frac{\partial^2 \cov}{\partial p_i\partial p_j}\invcov\cov\right]} \\
& \;\;\;\; - \Tr{\left[\invcov\frac{\partial \cov}{\partial p_i}\invcov\frac{\partial \cov}{\partial p_j}\invcov \cov\right]} + \Tr{\left[\invcov\frac{\partial\cov}{\partial p_j}\invcov\frac{\partial\cov}{\partial p_i} \right]}  - \Tr{\left[\invcov\frac{\partial^2\cov}{\partial p_i\partial p_j} \right]} \\
&= - 2\Tr\left[ \frac{\partial \model^T}{\partial p_i}\invcov \frac{\partial \model}{\partial p_j}\right] -\Tr{\left[\invcov\frac{\partial \cov}{\partial p_i}\invcov\frac{\partial \cov}{\partial p_j}\right]}.
\end{split}
\end{equation}
So therefore, the Fisher matrix for a Gaussian likelihood is given by
\begin{equation}\label{eq:generalFish}
    F_{ij} = \Tr\left[ \frac{\partial \model^T}{\partial p_i}\invcov \frac{\partial \model}{\partial p_j}\right] + \frac{1}{2}\Tr\Big[\invcov\frac{\partial \cov}{\partial p_i}\invcov\frac{\partial \mathbf{\Sigma}}{\partial p_j}\Big] .
\end{equation}

This Fisher matrix formalism that we have derived allows you to account for statistical errors that cause a widening of your error bars. However, it does not account for systematic errors due to inaccuracies in the model. These small residual systematics can bias the fiducial value ($p_i^\mathrm{fid}$) by $p_i^\mathrm{sys}$ such that the measured cosmological parameters are \(p_i = p_i^\mathrm{fid} + p_i^\mathrm{sys}\). To estimate the effect of these systematics, we can use the Fisher bias. In order to derive the Fisher bias, we first assume that the systematic error is small and that it affects the measured data as \(d_i = d_i^\mathrm{fid} + d_i^\mathrm{sys}\). Let us define the likelihoods $\mathcal{L}_0=\mathcal{L}(\dat,\param)$ with a maximum at $\param_0$ and $\mathcal{L}_{\rm sys}=\mathcal{L}(\dat,\param)$ with a maximum at $\param_0 + \param_{\rm sys}$. The former can be written out as in Eq(\ref{eq:loglike}) and the latter is
\begin{equation}
\begin{split}
    \loglike(\dat|\param)_{\rm sys} &= -\tfrac{1}{2}[\dat-\model+\dat^{\rm sys}]^T\invcov[\dat-\model+\dat^{\rm sys}] -\tfrac{1}{2}\ln{(2\pi)} -\tfrac{1}{2}\ln{[\det{\cov}]} \\
    &= -2\loglike_0 + 2(\dat^{\rm sys})^T\invcov(\dat-\model)-(\dat^{\rm sys})^T\invcov\dat^{\rm sys}.
\end{split}
\end{equation}
To find an expression for $\param_{\rm sys}$, we start by taking the maximum of $\loglike_{\rm sys}$:
\begin{equation}
\begin{split}
    0 &= -2\left.\left\langle\frac{\partial\loglike_{\rm sys}}{\partial p_j}\right\rangle\right|_{\param=\param_0+\param_{\rm sys}} \\
    &= \left.\left[ 
    \frac{\partial}{\partial p_j}(-2 \left\langle\loglike_0\right\rangle) + 2\left\langle(\dat^{\rm sys})^T\left\{ \frac{\partial \invcov}{\partial p_j}(\dat-\model) - \invcov\frac{\partial \model}{\partial p_j} \right\}\right\rangle + \left\langle(\dat^{\rm sys})^T\frac{\partial \invcov}{\partial p_j}\dat^{\rm sys}\right\rangle \right]\right|_{\param=\param_0+\param_{\rm sys}}.
\end{split}
\end{equation}
For the first term, the Fisher approximation that the likelihood is Gaussian in \param can be applied (Eq(\ref{eq:fishertaylor})) along with the definition of the Fisher matrix (Eq(\ref{eq:Fij}))
\begin{equation}
\begin{split}
    0 &= 2\sum_k F_{jk}p_k^{\rm sys} - 2\left\langle(\dat^{\rm sys})^T\left\{ \invcov \frac{\partial \cov}{\partial p_j}\invcov (\dat-\model) + \invcov\frac{\partial \model}{\partial p_j} \right\}\right\rangle - \left\langle(\dat^{\rm sys})^T\invcov \frac{\partial \cov}{\partial p_j}\invcov \dat^{\rm sys}\right\rangle \\
    &= 2\sum_k F_{jk}p_k^{\rm sys} - 2\left\langle(\dat^{\rm sys})^T\left\{ \invcov \frac{\partial \cov}{\partial p_j}\invcov (\dat-\model) + \invcov\frac{\partial \model}{\partial p_j} \right\}\right\rangle - \left\langle(\dat^{\rm sys})^T\invcov \frac{\partial \cov}{\partial p_j}\invcov \dat^{\rm sys}\right\rangle. 
\end{split}
\end{equation}
Then, we can multiply by $(F^{-1})_{ij}$ and solve for $\param_{\rm sys}$ to find that the Fisher bias is
\begin{align}\label{eq:generalBias}
    p_i^\mathrm{sys} &= \sum_j (F^{-1})_{ij} \Bigg\{ \mathrm{tr}\left[\left\langle(\mathbf{d}^\mathrm{sys})^\mathrm{T} \mathbf{\Sigma}^{-1} \frac{\partial \mathbf{\Sigma}}{\partial p_j} \mathbf{\Sigma}^{-1} (\mathbf{d}-\mathbf{m}+\frac{1}{2}\mathbf{d}^\mathrm{sys})\right\rangle \right] + \mathrm{tr}\left[\left\langle(\mathbf{d}^\mathrm{sys})^\mathrm{T} \mathbf{\Sigma}^{-1} \frac{\partial \mathbf{m}}{\partial p_j} \right\rangle\right] \Bigg\}. 
\end{align}

\end{widetext}

\subsection{Field Perspective}\label{sec:field}
In the field picture, our observables are the spherical harmonics $\dat=a_{\ell m}$ of density fluctuations, with $N_d=\sum_{\ell_{\rm min}}^{\ell_{\rm max}} (2\ell+1)$, and an expectation value $\model=\langle a_{\ell m}\rangle =0$ independent of cosmology. All cosmology dependence enters through the covariance, which is given by Eq(\ref{eq:fishercovariance}). 

If we apply this perspective to the general Fisher matrix equation we derived, Eq(\ref{eq:generalFish}), then the trace and matrix multiplications will be over all spherical harmonic $(\ell,m)$ components, resulting in
\begin{equation}\label{eq:Fmatrixfield}
    F^{\mathrm{field}}_{ij} = \sum_{\ell=\ell_\mathrm{min}}^{\ell_\mathrm{max}} \frac{(2\ell+1)f_\mathrm{sky}}{2} \mathrm{tr}\Big[C_\ell^{-1}\frac{\partial C_\ell}{\partial p_i}C_\ell^{-1}\frac{\partial C_\ell}{\partial p_j}\Big].
\end{equation}
and Fisher bias
\begin{align}\label{eq:Fbiasfield}
\begin{split}
    &p_i^\mathrm{field, sys} = \\
    &\sum_j (F^{-1})_{ij} \sum_{\ell=\ell_\mathrm{min}}^{\ell_\mathrm{max}} \frac{(2\ell+1)f_\mathrm{sky}}{2}  \mathrm{tr}\left[  C_\ell^{-1}\frac{\partial C_\ell}{\partial p_j}C_\ell^{-1} C_\ell^\mathrm{sys} \right].\\
\end{split}
\end{align}
Here we have defined 
\begin{equation}
\begin{split}
    C_\ell^\mathrm{sys}&=2\delta_{\ell \ell'}\delta_{mm'}\left\langle a_{\ell m}(a_{\ell' m'}^{\rm sys})^T\right\rangle\\ 
    & + \delta_{\ell \ell'}\delta_{mm'}\left\langle a_{\ell m}^{\rm sys}(a_{\ell' m'}^{\rm sys})^T\right\rangle\\
\end{split}
\end{equation}
which can also be written as the difference between the observed and the fiducial power spectra \(C_\ell^\mathrm{sys}=C_\ell^\mathrm{obs}-C_\ell^\mathrm{fid}\).

For the case of multiple tracers, the data is the collection of all $\dat=a^x_{\ell m}$'s of various tracers $x$, with $N_d=N_{\rm tracers}\sum_{\ell=1}^{\ell_{\rm max}} (2\ell+1)$ and the model is $\model=\langle a^x_{\ell m}\rangle =0$. The covariance elements are built from the angular power spectra $C^{xy}_{\ell}$, where for cross-correlations between tracers $x$ and $y$, 
\begin{equation}
   \cov_{(x,\ell,m)(y,\ell',m')}  = \langle a^x_{\ell m}(a^y_{\ell' m'})^*\rangle = \delta_{\ell \ell'}\delta_{m m'}C^{xy}_{\ell}.
\end{equation}
This changes the Fisher matrix expression by making the trace and matrix multiplications be over all possible tracer combinations, as well as, all spherical harmonic $(\ell,m)$ components. Since there is no dependence on $m$ in the covariance term, and the covariance is diagonal in $\ell$, we can simplify this by defining \cov as block diagonal, with $N_{\rm tracer}\times N_{\rm tracer}$ submatrices $\mathcal{C}_{\ell}$:
\begin{equation}\label{eq:fieldfisher_sph}
    F^{\mathrm{field}}_{ij} =  \sum_{\ell=\ell_\mathrm{min}}^{\ell_\mathrm{max}} \frac{(2\ell+1)f_\mathrm{sky}}{2} \Tr{\left[\mathcal{C}_{\ell}^{-1}\frac{\partial\mathcal{C}_{\ell}}{\partial p_j}\mathcal{C}_{\ell}^{-1}\frac{\partial \mathcal{C}_{\ell}}{\partial p_i}\right]}.
\end{equation}
In this expression, matrix operations are over different tracer combinations, and 
\begin{equation}
[\mathcal{C}_{\ell}]_{xy} = C^{xy}_{\ell}.
\end{equation}
Similarly, we can obtain the Fisher bias expression of 
\begin{align}\label{eq:fieldfisherbias_sph}
\begin{split}
    &p_i^\mathrm{field, sys}= \\
    & \sum_j (F^{-1})_{ij} \sum_{\ell=\ell_\mathrm{min}}^{\ell_\mathrm{max}} \frac{(2\ell+1)f_\mathrm{sky}}{2} \mathrm{tr}\left[  \mathcal{C}_\ell^{-1}\frac{\partial \mathcal{C}_\ell}{\partial p_j}\mathcal{C}_\ell^{-1} \mathcal{C}_\ell^\mathrm{sys} \right].
\end{split}
\end{align}

\subsection{Estimator Perspective}\label{sec:est}
Alternatively, the estimator perspective can be taken where the observables are the angular power spectra $\mathbf{d}=\hat{C_\ell}$ over some range of $\ell_\mathrm{min}-\ell_\mathrm{max}$ and $\mathbf{m}=\langle\hat{C_\ell}\rangle=C_\ell$. It can be shown using Wick's theorem that the covariance matrix is then given by 
\[\Sigma_{\ell\ell'}=\langle\hat{C_\ell}\hat{C_\ell'}\rangle-\langle\hat{C_\ell}\rangle\langle\hat{C_\ell'}\rangle=\delta_{\ell\ell'}\frac{2C_\ell^2}{(2\ell+1)}.\] This gives the Fisher matrix
\begin{align}\label{eq:Fmatrixest}
\begin{split}
    F^{\mathrm{est}}_{ij} &= f_\mathrm{sky}\Bigg\{ \sum_{\ell=\ell_\mathrm{min}}^{\ell_\mathrm{max}} \frac{(2\ell+1)}{2} \mathrm{tr}\Big[C_\ell^{-1}\frac{\partial C_\ell}{\partial p_i}C_\ell^{-1}\frac{\partial C_\ell}{\partial p_j}\Big] \\
    & + 2 \sum_{\ell=\ell_\mathrm{min}}^{\ell_\mathrm{max}} \mathrm{tr} \left[C_\ell^{-1}\frac{\partial C_\ell}{\partial p_i}C_\ell^{-1}\frac{\partial C_\ell}{\partial p_j}\right] \Bigg\} \\
\end{split}
\end{align}
where the second term is neglected in order to match with the field perspective. Since the second term comes from the derivative of the covariance matrix with respect to model parameters, it is often stated that this can be ignored by taking the covariance matrix to be parameter independent. However, it was shown by Carron~\cite{Carron:2013} that the actual reason we can neglect the second term of Eq(\ref{eq:Fmatrixest}) is because the $\hat{C_\ell}$ estimator is not in fact Gaussian, it is a gamma distribution. So going through the Fisher matrix derivation using the correct gamma distribution will in fact give the same Fisher matrix result for the estimator perspective as we found for the field perspective. This logic can similarly be used to derive the Fisher bias expression which is also the same as the field perspective. 

If we now extend this to multiple tracers, then the data becomes $\dat=\hat{C_\ell}^{xy}$ with expectation value $\mathbf{m}=\langle\hat{C_\ell}^{xy}\rangle=C_\ell^{xy}$. This means the data vector has length $N_{d} = N_{\rm tracer\, pairs}\times N_{\ell}$. Employing Wick's theorem again, one can derive that the covariance matrix for Gaussian fluctuations is given by 
\begin{equation}
    \cov_{(x,y,\ell)(u,v,\ell')} = \frac{\delta_{\ell\ell'}}{f_{\rm sky}(2\ell+1)}\left[C_{\ell}^{xu}C_{\ell}^{yv} + C_{\ell}^{xv}C_{\ell}^{yu}\right].
\end{equation}

It is convenient to not substitute out \model and just let it represent the vector of auto- and cross-correlations of tracers you are including into your Fisher matrix, such that
\begin{equation}\label{eq:estfisher_sph}
F_{ij} = \sum_{a=1}^{N_d} \sum_{b=1}^{N_d} \frac{\partial m_a}{\partial p_i} \Sigma_{ab}^{-1} \frac{\partial m_b}{\partial p_j} 
\end{equation}
and Fisher bias is 
\begin{equation}\label{eq:estfisherbias_sph}
     p_i^\mathrm{sys} = \sum_j (F^{-1})_{ij}\, \sum_{a=1}^{N_d} \sum_{b=1}^{N_d} m_a^\mathrm{sys}  \Sigma_{ab}^{-1} \frac{\partial m_b}{\partial p_j}.
\end{equation}

\subsection{Discussion of Field-Estimator Equivalence}\label{sec:fieldest}

When performing Fisher forecasts in cosmology, there are some subtleties that arise depending on what you define as your observable (or which \textit{perspective} you use). 
The field and estimator perspectives are equivalent for a single tracer or for multiple tracers with all auto- and cross-correlations included. However, if there are multiple tracers, but only a subset of correlations are considered as the observable (ie. only auto-correlations), then these two approaches will produce slightly different results (to the extent that both are even possible to appropriately define). 

To understand how this difference between perspectives arises, let's consider changing the Fisher matrix equation from using auto+cross-bin correlations to only auto-bin correlations. For the estimator perspective, to go from the auto+cross-bin equation, Eq(\ref{eq:estfisher_sph}), to auto-bin only, we just need to change the model vector. When $\model$ is input, it would only include the auto-correlations since that is the only data being studied. The covariance matrix would still have both auto- and cross-correlations because the theory should account for the presence of cross-correlations even if our data does not measure them. This is equivalent to marginalizing over the spectra that are not included in the data vector used in the analysis. Turning to the field perspective, Eq(\ref{eq:fieldfisher_sph}), the $1/\mathcal{C}_{\ell}$ terms would also contain the auto- and cross-correlations because they arise from the covariance in the derivation. However, the $\partial \mathcal{C}_{\ell}/\partial p_i$ terms inform us about how sensitive our observables are to the parameter $p_i$, so if cross-correlations are not included as an observable the likelihood analysis should be insensitive to their reactions to these parameters. Therefore, the $\partial \mathcal{C}_\ell/\partial p_i$ terms with cross-correlations are set to zero. It is not clear in the field picture what to marginalize over in the likelihood to produce a corresponding result. To summarize the difference between the estimator and field approaches, in the former the $\model$ neglects some correlations, whereas the latter sets a subset of derivatives to zero. 

Applying these two methodologies to our analysis, we found that by estimating constraining power using the field perspective, the Fisher matrix is slightly overestimated in comparison to the estimator result. Thus, the field approach will systematically underestimate parameter error bars and overestimate bias. We can see this for instance in the constraints we obtained on $\epsilon$, shown in Table~\ref{table:DESY3estvsfield}. This distinction between perspectives is very relevant to note when comparing forecast results, or trying to implement a forecast which is an accurate mock of a survey collaboration's data analysis.

\begin{table}
\centering
\begin{tabular}{l|c|c}
    Bin Combinations & $\sigma(\epsilon)$ Field  & $\sigma(\epsilon)$ Estimator \\
   \hline 
   \hline
     Bin 1$\times$Bin 1 & 5.08 & 5.08 \\
    \hline
     Bin 2$\times$Bin 2 &  2.51 &  2.51 \\
    \hline
     Bin 3$\times$Bin 3 & 0.92 & 0.92 \\
     \hline 
     Bin 4$\times$Bin 4 & 0.77 & 0.77 \\
     \hline 
     \textbf{All Auto-Bins} & \textbf{0.52} & \textbf{0.59} \\
     \hline 
     All Auto+Cross-Bins & 0.28 &  0.28 \\
\end{tabular}
\captionsetup{justification=raggedright}
\caption{Constraints on $\epsilon$ for DES Y3, obtained from both the field and estimator perspectives. The results only differ for the case of auto-bin correlations only (highlighted in bold), where the field perspective constraint is systematically smaller than the estimator constraint.}
\label{table:DESY3estvsfield}
\end{table}

\section{Numerical Derivatives} \label{app:numderiv}

We forecast constraints the cosmological parameters $\Omega_{\rm c}$ and $\sigma_8$, along with the parameters $b_g^i$, $\Delta z^i$, and $\sigma_{z}^i$ for each sample of galaxies, fixing all other cosmological parameters to their fiducial values.  
The angular power spectra's linear or quadratic dependence on $b_g^i$ allow us to use exact derivatives with respect to galaxy bias, while other parameters require numerical derivatives. For parameters $[\Omega_{\rm c}, \sigma_8, \Delta z^i, \sigma_{z}^i]$ we compute numerical derivatives using a forward finite difference approach as follows.
For the derivative of quantity $f$ with respect to a parameter evaluated at value $x$, and for a step size $h$ describing a small change in the parameter value,
\begin{equation}
    f'(x) \approx \frac{f(x+h)-f(x)}{h}.
\end{equation}
This finite difference approach is accurate up to order $h$. For $\Omega_{\rm c}$ and $\sigma_8$, we compute this numerical derivative directly using calculations of angular power spectra $C_{\ell}$'s. For the redshift nuisance parameters $\Delta z^i$ and $\sigma_{z}^i$, which affect power spectra only through their impact on $n(z)$, we find more stable behavior by computing numerical derivatives for the $n(z)$ distributions, and then inputting those derivatives as galaxy window functions into \code{CAMB} $C_{\ell}$ calculations. 
Note that this takes some care to account for the fact that  \code{CAMB} automatically normalizes and takes the absolute value of input galaxy window functions. 

If we denote the numerical derivative as $\mathbf{D}(x,h)$, then an $\mathcal{O}(h)$ derivative should converge as
\begin{equation}
     \frac{\lvert\lvert \mathbf{D}(x,h) - \mathbf{D}(x,h/2)\rvert\rvert_2}{\lvert\lvert \mathbf{D}(x,h/2) - \mathbf{D}(x,h/4)\rvert\rvert_2} = 2.
\end{equation}
We use this convergence check to ensure that our derivatives followed the expected error. For the DES cosmology parameters, the auto- and cross-bin convergence rates spanned (1.43-2.37) for $\Omega_{\rm c}$ and were all 1.998 for $\sigma_8$. LSST, had the $\Omega_{\rm c}$ convergence rate span (1.86-2.03), and for $\sigma_8$ we achieved 1.9998.

Additionally we study the impact of changing derivative step size on the unmarginalized Fisher error computed via \eqs{eq:fisherdef} and~\ref{eq:unmargerr}. 
This is a useful metric because it allows us to check the robustness of a quantity we are most concerned about for our forecasts. For each parameter we check the percent difference change in $\sigma(p_i)$ introduced by changing the derivative step size from $h$ to $2h$. For both DES and LSST forecast, we find this to be under 10\% for all parameters, and below 1\% for all cosmological parameters.



\bibliography{apssamp}

\providecommand{\noopsort}[1]{}\providecommand{\singleletter}[1]{#1}%
\begin{thebibliography}{72}%
\makeatletter
\providecommand \@ifxundefined [1]{%
 \@ifx{#1\undefined}
}%
\providecommand \@ifnum [1]{%
 \ifnum #1\expandafter \@firstoftwo
 \else \expandafter \@secondoftwo
 \fi
}%
\providecommand \@ifx [1]{%
 \ifx #1\expandafter \@firstoftwo
 \else \expandafter \@secondoftwo
 \fi
}%
\providecommand \natexlab [1]{#1}%
\providecommand \enquote  [1]{``#1''}%
\providecommand \bibnamefont  [1]{#1}%
\providecommand \bibfnamefont [1]{#1}%
\providecommand \citenamefont [1]{#1}%
\providecommand \href@noop [0]{\@secondoftwo}%
\providecommand \href [0]{\begingroup \@sanitize@url \@href}%
\providecommand \@href[1]{\@@startlink{#1}\@@href}%
\providecommand \@@href[1]{\endgroup#1\@@endlink}%
\providecommand \@sanitize@url [0]{\catcode `\\12\catcode `\$12\catcode
  `\&12\catcode `\#12\catcode `\^12\catcode `\_12\catcode `\%12\relax}%
\providecommand \@@startlink[1]{}%
\providecommand \@@endlink[0]{}%
\providecommand \url  [0]{\begingroup\@sanitize@url \@url }%
\providecommand \@url [1]{\endgroup\@href {#1}{\urlprefix }}%
\providecommand \urlprefix  [0]{URL }%
\providecommand \Eprint [0]{\href }%
\providecommand \doibase [0]{https://doi.org/}%
\providecommand \selectlanguage [0]{\@gobble}%
\providecommand \bibinfo  [0]{\@secondoftwo}%
\providecommand \bibfield  [0]{\@secondoftwo}%
\providecommand \translation [1]{[#1]}%
\providecommand \BibitemOpen [0]{}%
\providecommand \bibitemStop [0]{}%
\providecommand \bibitemNoStop [0]{.\EOS\space}%
\providecommand \EOS [0]{\spacefactor3000\relax}%
\providecommand \BibitemShut  [1]{\csname bibitem#1\endcsname}%
\let\auto@bib@innerbib\@empty
\bibitem [{\citenamefont {{Flaugher}}\ \emph {et~al.}(2015)\citenamefont
  {{Flaugher}}, \citenamefont {{Diehl}}, \citenamefont {{Honscheid}} \emph
  {et~al.}}]{flaugher15}%
  \BibitemOpen
  \bibfield  {author} {\bibinfo {author} {\bibfnamefont {B.}~\bibnamefont
  {{Flaugher}}}, \bibinfo {author} {\bibfnamefont {H.~T.}\ \bibnamefont
  {{Diehl}}}, \bibinfo {author} {\bibfnamefont {K.}~\bibnamefont
  {{Honscheid}}}, \emph {et~al.} (\bibinfo {collaboration} {{DES}}),\ }\href
  {https://doi.org/10.1088/0004-6256/150/5/150} {\bibfield  {journal} {\bibinfo
   {journal} {The Astronomical Journal}\ }\textbf {\bibinfo {volume} {150}},\
  \bibinfo {eid} {150} (\bibinfo {year} {2015})},\ \Eprint
  {https://arxiv.org/abs/1504.02900} {arXiv:1504.02900 [astro-ph.IM]}
  \BibitemShut {NoStop}%
\bibitem [{\citenamefont {{DES Collaboration}}(2022{\natexlab{a}})}]{DES:Y3}%
  \BibitemOpen
  \bibfield  {author} {\bibinfo {author} {\bibnamefont {{DES Collaboration}}}
  (\bibinfo {collaboration} {DES}),\ }\href
  {https://doi.org/10.1103/PhysRevD.105.023520} {\bibfield  {journal} {\bibinfo
   {journal} {Phys. Rev. D}\ }\textbf {\bibinfo {volume} {105}},\ \bibinfo
  {pages} {023520} (\bibinfo {year} {2022}{\natexlab{a}})},\ \Eprint
  {https://arxiv.org/abs/2105.13549} {arXiv:2105.13549 [astro-ph.CO]}
  \BibitemShut {NoStop}%
\bibitem [{\citenamefont {{DES Collaboration}}(2018)}]{DES:Y1cosm}%
  \BibitemOpen
  \bibfield  {author} {\bibinfo {author} {\bibnamefont {{DES Collaboration}}}
  (\bibinfo {collaboration} {DES}),\ }\href
  {https://doi.org/10.1103/PhysRevD.98.043526} {\bibfield  {journal} {\bibinfo
  {journal} {Phys. Rev. D}\ }\textbf {\bibinfo {volume} {98}},\ \bibinfo
  {pages} {043526} (\bibinfo {year} {2018})},\ \Eprint
  {https://arxiv.org/abs/1708.01530} {arXiv:1708.01530 [astro-ph.CO]}
  \BibitemShut {NoStop}%
\bibitem [{\citenamefont {Heymans}\ \emph {et~al.}(2021)\citenamefont {Heymans}
  \emph {et~al.}}]{Heymans:2020gsg}%
  \BibitemOpen
  \bibfield  {author} {\bibinfo {author} {\bibfnamefont {C.}~\bibnamefont
  {Heymans}} \emph {et~al.},\ }\href
  {https://doi.org/10.1051/0004-6361/202039063} {\bibfield  {journal} {\bibinfo
   {journal} {Astron. Astrophys.}\ }\textbf {\bibinfo {volume} {646}},\
  \bibinfo {pages} {A140} (\bibinfo {year} {2021})},\ \Eprint
  {https://arxiv.org/abs/2007.15632} {arXiv:2007.15632 [astro-ph.CO]}
  \BibitemShut {NoStop}%
\bibitem [{\citenamefont {Dvornik}\ \emph {et~al.}(2022)\citenamefont {Dvornik}
  \emph {et~al.}}]{Dvornik:2022xap}%
  \BibitemOpen
  \bibfield  {author} {\bibinfo {author} {\bibfnamefont {A.}~\bibnamefont
  {Dvornik}} \emph {et~al.}\ }\href
  {https://doi.org/10.1051/0004-6361/202245158} {10.1051/0004-6361/202245158}
  (\bibinfo {year} {2022}),\ \Eprint {https://arxiv.org/abs/2210.03110}
  {arXiv:2210.03110 [astro-ph.CO]} \BibitemShut {NoStop}%
\bibitem [{\citenamefont {More}\ \emph {et~al.}(2023)\citenamefont {More} \emph
  {et~al.}}]{More:2023knf}%
  \BibitemOpen
  \bibfield  {author} {\bibinfo {author} {\bibfnamefont {S.}~\bibnamefont
  {More}} \emph {et~al.},\ }\href@noop {} {\  (\bibinfo {year} {2023})},\
  \Eprint {https://arxiv.org/abs/2304.00703} {arXiv:2304.00703 [astro-ph.CO]}
  \BibitemShut {NoStop}%
\bibitem [{\citenamefont {Sugiyama}\ \emph {et~al.}(2023)\citenamefont
  {Sugiyama} \emph {et~al.}}]{Sugiyama:2023fzm}%
  \BibitemOpen
  \bibfield  {author} {\bibinfo {author} {\bibfnamefont {S.}~\bibnamefont
  {Sugiyama}} \emph {et~al.},\ }\href@noop {} {\  (\bibinfo {year} {2023})},\
  \Eprint {https://arxiv.org/abs/2304.00705} {arXiv:2304.00705 [astro-ph.CO]}
  \BibitemShut {NoStop}%
\bibitem [{\citenamefont {Miyatake}\ \emph {et~al.}(2023)\citenamefont
  {Miyatake} \emph {et~al.}}]{Miyatake:2023njf}%
  \BibitemOpen
  \bibfield  {author} {\bibinfo {author} {\bibfnamefont {H.}~\bibnamefont
  {Miyatake}} \emph {et~al.},\ }\href@noop {} {\  (\bibinfo {year} {2023})},\
  \Eprint {https://arxiv.org/abs/2304.00704} {arXiv:2304.00704 [astro-ph.CO]}
  \BibitemShut {NoStop}%
\bibitem [{\citenamefont {Vakili}\ \emph {et~al.}(2023)\citenamefont {Vakili}
  \emph {et~al.}}]{Vakili:2020dwl}%
  \BibitemOpen
  \bibfield  {author} {\bibinfo {author} {\bibfnamefont {M.}~\bibnamefont
  {Vakili}} \emph {et~al.},\ }\href
  {https://doi.org/10.1051/0004-6361/202039293} {\bibfield  {journal} {\bibinfo
   {journal} {Astron. Astrophys.}\ }\textbf {\bibinfo {volume} {675}},\
  \bibinfo {pages} {A202} (\bibinfo {year} {2023})},\ \Eprint
  {https://arxiv.org/abs/2008.13154} {arXiv:2008.13154 [astro-ph.CO]}
  \BibitemShut {NoStop}%
\bibitem [{\citenamefont {Nicola}\ \emph {et~al.}(2020)\citenamefont {Nicola}
  \emph {et~al.}}]{Nicola:HSCclustering}%
  \BibitemOpen
  \bibfield  {author} {\bibinfo {author} {\bibfnamefont {A.}~\bibnamefont
  {Nicola}} \emph {et~al.} (\bibinfo {collaboration} {LSST}),\ }\href
  {https://doi.org/10.1088/1475-7516/2020/03/044} {\bibfield  {journal}
  {\bibinfo  {journal} {JCAP}\ }\textbf {\bibinfo {volume} {03}},\ \bibinfo
  {pages} {044}},\ \Eprint {https://arxiv.org/abs/1912.08209} {arXiv:1912.08209
  [astro-ph.CO]} \BibitemShut {NoStop}%
\bibitem [{\citenamefont {Xu}\ \emph {et~al.}(2023)\citenamefont {Xu} \emph
  {et~al.}}]{Xu:2023zlv}%
  \BibitemOpen
  \bibfield  {author} {\bibinfo {author} {\bibfnamefont {H.}~\bibnamefont {Xu}}
  \emph {et~al.},\ }\href@noop {} {\  (\bibinfo {year} {2023})},\ \Eprint
  {https://arxiv.org/abs/2310.03066} {arXiv:2310.03066 [astro-ph.CO]}
  \BibitemShut {NoStop}%
\bibitem [{\citenamefont {Coupon}\ \emph {et~al.}(2012)\citenamefont {Coupon}
  \emph {et~al.}}]{Coupon:2011mz}%
  \BibitemOpen
  \bibfield  {author} {\bibinfo {author} {\bibfnamefont {J.}~\bibnamefont
  {Coupon}} \emph {et~al.},\ }\href
  {https://doi.org/10.1051/0004-6361/201117625} {\bibfield  {journal} {\bibinfo
   {journal} {Astron. Astrophys.}\ }\textbf {\bibinfo {volume} {542}},\
  \bibinfo {pages} {A5} (\bibinfo {year} {2012})},\ \Eprint
  {https://arxiv.org/abs/1107.0616} {arXiv:1107.0616 [astro-ph.CO]}
  \BibitemShut {NoStop}%
\bibitem [{\citenamefont {Padmanabhan}\ \emph {et~al.}(2007)\citenamefont
  {Padmanabhan} \emph {et~al.}}]{Padmanabhan:2006egz}%
  \BibitemOpen
  \bibfield  {author} {\bibinfo {author} {\bibfnamefont {N.}~\bibnamefont
  {Padmanabhan}} \emph {et~al.} (\bibinfo {collaboration} {SDSS}),\ }\href
  {https://doi.org/10.1111/j.1365-2966.2007.11593.x} {\bibfield  {journal}
  {\bibinfo  {journal} {Mon. Not. Roy. Astron. Soc.}\ }\textbf {\bibinfo
  {volume} {378}},\ \bibinfo {pages} {852} (\bibinfo {year} {2007})},\ \Eprint
  {https://arxiv.org/abs/astro-ph/0605302} {arXiv:astro-ph/0605302}
  \BibitemShut {NoStop}%
\bibitem [{\citenamefont {Blake}\ \emph {et~al.}(2007)\citenamefont {Blake},
  \citenamefont {Collister}, \citenamefont {Bridle},\ and\ \citenamefont
  {Lahav}}]{Blake:2006kv}%
  \BibitemOpen
  \bibfield  {author} {\bibinfo {author} {\bibfnamefont {C.}~\bibnamefont
  {Blake}}, \bibinfo {author} {\bibfnamefont {A.}~\bibnamefont {Collister}},
  \bibinfo {author} {\bibfnamefont {S.}~\bibnamefont {Bridle}},\ and\ \bibinfo
  {author} {\bibfnamefont {O.}~\bibnamefont {Lahav}},\ }\href
  {https://doi.org/10.1111/j.1365-2966.2006.11263.x} {\bibfield  {journal}
  {\bibinfo  {journal} {Mon. Not. Roy. Astron. Soc.}\ }\textbf {\bibinfo
  {volume} {374}},\ \bibinfo {pages} {1527} (\bibinfo {year} {2007})},\ \Eprint
  {https://arxiv.org/abs/astro-ph/0605303} {arXiv:astro-ph/0605303}
  \BibitemShut {NoStop}%
\bibitem [{\citenamefont {Estrada}\ \emph {et~al.}(2009)\citenamefont
  {Estrada}, \citenamefont {Sefusatti},\ and\ \citenamefont
  {Frieman}}]{Estrada:2008em}%
  \BibitemOpen
  \bibfield  {author} {\bibinfo {author} {\bibfnamefont {J.}~\bibnamefont
  {Estrada}}, \bibinfo {author} {\bibfnamefont {E.}~\bibnamefont {Sefusatti}},\
  and\ \bibinfo {author} {\bibfnamefont {J.~A.}\ \bibnamefont {Frieman}},\
  }\href {https://doi.org/10.1088/0004-637X/692/1/265} {\bibfield  {journal}
  {\bibinfo  {journal} {Astrophys. J.}\ }\textbf {\bibinfo {volume} {692}},\
  \bibinfo {pages} {265} (\bibinfo {year} {2009})},\ \Eprint
  {https://arxiv.org/abs/0801.3485} {arXiv:0801.3485 [astro-ph]} \BibitemShut
  {NoStop}%
\bibitem [{\citenamefont {de~Putter}\ \emph {et~al.}(2012)\citenamefont
  {de~Putter} \emph {et~al.}}]{dePutter:2012sh}%
  \BibitemOpen
  \bibfield  {author} {\bibinfo {author} {\bibfnamefont {R.}~\bibnamefont
  {de~Putter}} \emph {et~al.},\ }\href
  {https://doi.org/10.1088/0004-637X/761/1/12} {\bibfield  {journal} {\bibinfo
  {journal} {Astrophys. J.}\ }\textbf {\bibinfo {volume} {761}},\ \bibinfo
  {pages} {12} (\bibinfo {year} {2012})},\ \Eprint
  {https://arxiv.org/abs/1201.1909} {arXiv:1201.1909 [astro-ph.CO]}
  \BibitemShut {NoStop}%
\bibitem [{\citenamefont {Ho}\ \emph {et~al.}(2015)\citenamefont {Ho} \emph
  {et~al.}}]{Ho:2013lda}%
  \BibitemOpen
  \bibfield  {author} {\bibinfo {author} {\bibfnamefont {S.}~\bibnamefont {Ho}}
  \emph {et~al.},\ }\href {https://doi.org/10.1088/1475-7516/2015/05/040}
  {\bibfield  {journal} {\bibinfo  {journal} {JCAP}\ }\textbf {\bibinfo
  {volume} {05}},\ \bibinfo {pages} {040}},\ \Eprint
  {https://arxiv.org/abs/1311.2597} {arXiv:1311.2597 [astro-ph.CO]}
  \BibitemShut {NoStop}%
\bibitem [{\citenamefont {Jouvel}\ \emph {et~al.}(2017)\citenamefont {Jouvel}
  \emph {et~al.}}]{DES:2015vbk}%
  \BibitemOpen
  \bibfield  {author} {\bibinfo {author} {\bibfnamefont {S.}~\bibnamefont
  {Jouvel}} \emph {et~al.} (\bibinfo {collaboration} {DES, BOSS}),\ }\href
  {https://doi.org/10.1093/mnras/stx163} {\bibfield  {journal} {\bibinfo
  {journal} {Mon. Not. Roy. Astron. Soc.}\ }\textbf {\bibinfo {volume} {469}},\
  \bibinfo {pages} {2771} (\bibinfo {year} {2017})},\ \Eprint
  {https://arxiv.org/abs/1509.07121} {arXiv:1509.07121 [astro-ph.CO]}
  \BibitemShut {NoStop}%
\bibitem [{\citenamefont {Balaguera-Antol\'\i{}nez}\ \emph
  {et~al.}(2018)\citenamefont {Balaguera-Antol\'\i{}nez}, \citenamefont
  {Bilicki}, \citenamefont {Branchini},\ and\ \citenamefont
  {Postiglione}}]{Balaguera-Antolinez:2017dpm}%
  \BibitemOpen
  \bibfield  {author} {\bibinfo {author} {\bibfnamefont {A.}~\bibnamefont
  {Balaguera-Antol\'\i{}nez}}, \bibinfo {author} {\bibfnamefont
  {M.}~\bibnamefont {Bilicki}}, \bibinfo {author} {\bibfnamefont
  {E.}~\bibnamefont {Branchini}},\ and\ \bibinfo {author} {\bibfnamefont
  {A.}~\bibnamefont {Postiglione}},\ }\href
  {https://doi.org/10.1093/mnras/sty262} {\bibfield  {journal} {\bibinfo
  {journal} {Mon. Not. Roy. Astron. Soc.}\ }\textbf {\bibinfo {volume} {476}},\
  \bibinfo {pages} {1050} (\bibinfo {year} {2018})},\ \Eprint
  {https://arxiv.org/abs/1711.04583} {arXiv:1711.04583 [astro-ph.CO]}
  \BibitemShut {NoStop}%
\bibitem [{\citenamefont {Blake}\ and\ \citenamefont
  {Wall}(2002)}]{Blake:2001bg}%
  \BibitemOpen
  \bibfield  {author} {\bibinfo {author} {\bibfnamefont {C.}~\bibnamefont
  {Blake}}\ and\ \bibinfo {author} {\bibfnamefont {J.}~\bibnamefont {Wall}},\
  }\href {https://doi.org/10.1046/j.1365-8711.2002.05163.x} {\bibfield
  {journal} {\bibinfo  {journal} {Mon. Not. Roy. Astron. Soc.}\ }\textbf
  {\bibinfo {volume} {329}},\ \bibinfo {pages} {L37} (\bibinfo {year}
  {2002})},\ \Eprint {https://arxiv.org/abs/astro-ph/0111328}
  {arXiv:astro-ph/0111328} \BibitemShut {NoStop}%
\bibitem [{\citenamefont {{DES Collaboration}}(2022{\natexlab{b}})}]{desy3bao}%
  \BibitemOpen
  \bibfield  {author} {\bibinfo {author} {\bibnamefont {{DES Collaboration}}}
  (\bibinfo {collaboration} {DES}),\ }\href
  {https://doi.org/10.1103/PhysRevD.105.043512} {\bibfield  {journal} {\bibinfo
   {journal} {Phys. Rev. D}\ }\textbf {\bibinfo {volume} {105}},\ \bibinfo
  {pages} {043512} (\bibinfo {year} {2022}{\natexlab{b}})},\ \Eprint
  {https://arxiv.org/abs/2107.04646} {arXiv:2107.04646 [astro-ph.CO]}
  \BibitemShut {NoStop}%
\bibitem [{\citenamefont {Seo}\ \emph {et~al.}(2012)\citenamefont {Seo} \emph
  {et~al.}}]{Seo:2012xy}%
  \BibitemOpen
  \bibfield  {author} {\bibinfo {author} {\bibfnamefont {H.-J.}\ \bibnamefont
  {Seo}} \emph {et~al.},\ }\href {https://doi.org/10.1088/0004-637X/761/1/13}
  {\bibfield  {journal} {\bibinfo  {journal} {Astrophys. J.}\ }\textbf
  {\bibinfo {volume} {761}},\ \bibinfo {pages} {13} (\bibinfo {year} {2012})},\
  \Eprint {https://arxiv.org/abs/1201.2172} {arXiv:1201.2172 [astro-ph.CO]}
  \BibitemShut {NoStop}%
\bibitem [{\citenamefont {Crocce}\ \emph {et~al.}(2011)\citenamefont {Crocce},
  \citenamefont {Gaztanaga}, \citenamefont {Cabre}, \citenamefont {Carnero},\
  and\ \citenamefont {Sanchez}}]{Crocce:2011mj}%
  \BibitemOpen
  \bibfield  {author} {\bibinfo {author} {\bibfnamefont {M.}~\bibnamefont
  {Crocce}}, \bibinfo {author} {\bibfnamefont {E.}~\bibnamefont {Gaztanaga}},
  \bibinfo {author} {\bibfnamefont {A.}~\bibnamefont {Cabre}}, \bibinfo
  {author} {\bibfnamefont {A.}~\bibnamefont {Carnero}},\ and\ \bibinfo {author}
  {\bibfnamefont {E.}~\bibnamefont {Sanchez}},\ }\href
  {https://doi.org/10.1111/j.1365-2966.2011.19425.x} {\bibfield  {journal}
  {\bibinfo  {journal} {Mon. Not. Roy. Astron. Soc.}\ }\textbf {\bibinfo
  {volume} {417}},\ \bibinfo {pages} {2577} (\bibinfo {year} {2011})},\ \Eprint
  {https://arxiv.org/abs/1104.5236} {arXiv:1104.5236 [astro-ph.CO]}
  \BibitemShut {NoStop}%
\bibitem [{\citenamefont {Carnero}\ \emph {et~al.}(2012)\citenamefont
  {Carnero}, \citenamefont {Sanchez}, \citenamefont {Crocce}, \citenamefont
  {Cabre},\ and\ \citenamefont {Gaztanaga}}]{Carnero:2011pu}%
  \BibitemOpen
  \bibfield  {author} {\bibinfo {author} {\bibfnamefont {A.}~\bibnamefont
  {Carnero}}, \bibinfo {author} {\bibfnamefont {E.}~\bibnamefont {Sanchez}},
  \bibinfo {author} {\bibfnamefont {M.}~\bibnamefont {Crocce}}, \bibinfo
  {author} {\bibfnamefont {A.}~\bibnamefont {Cabre}},\ and\ \bibinfo {author}
  {\bibfnamefont {E.}~\bibnamefont {Gaztanaga}},\ }\href
  {https://doi.org/10.1111/j.1365-2966.2011.19832.x} {\bibfield  {journal}
  {\bibinfo  {journal} {Mon. Not. Roy. Astron. Soc.}\ }\textbf {\bibinfo
  {volume} {419}},\ \bibinfo {pages} {1689} (\bibinfo {year} {2012})},\ \Eprint
  {https://arxiv.org/abs/1104.5426} {arXiv:1104.5426 [astro-ph.CO]}
  \BibitemShut {NoStop}%
\bibitem [{\citenamefont {Huetsi}(2010)}]{Huetsi:2009zq}%
  \BibitemOpen
  \bibfield  {author} {\bibinfo {author} {\bibfnamefont {G.}~\bibnamefont
  {Huetsi}},\ }\href {https://doi.org/10.1111/j.1365-2966.2009.15824.x}
  {\bibfield  {journal} {\bibinfo  {journal} {Mon. Not. Roy. Astron. Soc.}\
  }\textbf {\bibinfo {volume} {401}},\ \bibinfo {pages} {2477} (\bibinfo {year}
  {2010})},\ \Eprint {https://arxiv.org/abs/0910.0492} {arXiv:0910.0492
  [astro-ph.CO]} \BibitemShut {NoStop}%
\bibitem [{\citenamefont {Mandelbaum}\ \emph {et~al.}(2018)\citenamefont
  {Mandelbaum} \emph {et~al.}}]{LSST:DESC_SRD}%
  \BibitemOpen
  \bibfield  {author} {\bibinfo {author} {\bibfnamefont {R.}~\bibnamefont
  {Mandelbaum}} \emph {et~al.} (\bibinfo {collaboration} {LSST Dark Energy
  Science}),\ }\href@noop {} {\  (\bibinfo {year} {2018})},\ \Eprint
  {https://arxiv.org/abs/1809.01669} {arXiv:1809.01669 [astro-ph.CO]}
  \BibitemShut {NoStop}%
\bibitem [{\citenamefont {Laureijs}\ \emph {et~al.}(2011)\citenamefont
  {Laureijs} \emph {et~al.}}]{EUCLID:definitionreport}%
  \BibitemOpen
  \bibfield  {author} {\bibinfo {author} {\bibfnamefont {R.}~\bibnamefont
  {Laureijs}} \emph {et~al.} (\bibinfo {collaboration} {EUCLID}),\ }\href@noop
  {} {\  (\bibinfo {year} {2011})},\ \Eprint {https://arxiv.org/abs/1110.3193}
  {arXiv:1110.3193 [astro-ph.CO]} \BibitemShut {NoStop}%
\bibitem [{\citenamefont {Pocino}\ \emph {et~al.}(2021)\citenamefont {Pocino}
  \emph {et~al.}}]{Euclid:photoclustering_opt}%
  \BibitemOpen
  \bibfield  {author} {\bibinfo {author} {\bibfnamefont {A.}~\bibnamefont
  {Pocino}} \emph {et~al.} (\bibinfo {collaboration} {Euclid}),\ }\href
  {https://doi.org/10.1051/0004-6361/202141061} {\bibfield  {journal} {\bibinfo
   {journal} {Astron. Astrophys.}\ }\textbf {\bibinfo {volume} {655}},\
  \bibinfo {pages} {A44} (\bibinfo {year} {2021})},\ \Eprint
  {https://arxiv.org/abs/2104.05698} {arXiv:2104.05698 [astro-ph.CO]}
  \BibitemShut {NoStop}%
\bibitem [{\citenamefont {Spergel}\ \emph {et~al.}(2015)\citenamefont {Spergel}
  \emph {et~al.}}]{Spergel:2015sza}%
  \BibitemOpen
  \bibfield  {author} {\bibinfo {author} {\bibfnamefont {D.}~\bibnamefont
  {Spergel}} \emph {et~al.},\ }\href@noop {} {\  (\bibinfo {year} {2015})},\
  \Eprint {https://arxiv.org/abs/1503.03757} {arXiv:1503.03757 [astro-ph.IM]}
  \BibitemShut {NoStop}%
\bibitem [{\citenamefont {Eifler}\ \emph {et~al.}(2021)\citenamefont {Eifler}
  \emph {et~al.}}]{Eifler:2020vvg}%
  \BibitemOpen
  \bibfield  {author} {\bibinfo {author} {\bibfnamefont {T.}~\bibnamefont
  {Eifler}} \emph {et~al.},\ }\href {https://doi.org/10.1093/mnras/stab1762}
  {\bibfield  {journal} {\bibinfo  {journal} {Mon. Not. Roy. Astron. Soc.}\
  }\textbf {\bibinfo {volume} {507}},\ \bibinfo {pages} {1746} (\bibinfo {year}
  {2021})},\ \Eprint {https://arxiv.org/abs/2004.05271} {arXiv:2004.05271
  [astro-ph.CO]} \BibitemShut {NoStop}%
\bibitem [{\citenamefont {Porredon}\ \emph {et~al.}(2022)\citenamefont
  {Porredon} \emph {et~al.}}]{DES:maglim2x2pt_results}%
  \BibitemOpen
  \bibfield  {author} {\bibinfo {author} {\bibfnamefont {A.}~\bibnamefont
  {Porredon}} \emph {et~al.} (\bibinfo {collaboration} {DES}),\ }\href
  {https://doi.org/10.1103/PhysRevD.106.103530} {\bibfield  {journal} {\bibinfo
   {journal} {Phys. Rev. D}\ }\textbf {\bibinfo {volume} {106}},\ \bibinfo
  {pages} {103530} (\bibinfo {year} {2022})},\ \Eprint
  {https://arxiv.org/abs/2105.13546} {arXiv:2105.13546 [astro-ph.CO]}
  \BibitemShut {NoStop}%
\bibitem [{\citenamefont {Pandey}\ \emph {et~al.}(2022)\citenamefont {Pandey}
  \emph {et~al.}}]{DES:redmagic2x2pt_results}%
  \BibitemOpen
  \bibfield  {author} {\bibinfo {author} {\bibfnamefont {S.}~\bibnamefont
  {Pandey}} \emph {et~al.} (\bibinfo {collaboration} {DES}),\ }\href
  {https://doi.org/10.1103/PhysRevD.106.043520} {\bibfield  {journal} {\bibinfo
   {journal} {Phys. Rev. D}\ }\textbf {\bibinfo {volume} {106}},\ \bibinfo
  {pages} {043520} (\bibinfo {year} {2022})},\ \Eprint
  {https://arxiv.org/abs/2105.13545} {arXiv:2105.13545 [astro-ph.CO]}
  \BibitemShut {NoStop}%
\bibitem [{\citenamefont {Porredon}\ \emph {et~al.}(2021)\citenamefont
  {Porredon} \emph {et~al.}}]{DES:maglim_optimization}%
  \BibitemOpen
  \bibfield  {author} {\bibinfo {author} {\bibfnamefont {A.}~\bibnamefont
  {Porredon}} \emph {et~al.} (\bibinfo {collaboration} {DES}),\ }\href
  {https://doi.org/10.1103/PhysRevD.103.043503} {\bibfield  {journal} {\bibinfo
   {journal} {Phys. Rev. D}\ }\textbf {\bibinfo {volume} {103}},\ \bibinfo
  {pages} {043503} (\bibinfo {year} {2021})},\ \Eprint
  {https://arxiv.org/abs/2011.03411} {arXiv:2011.03411 [astro-ph.CO]}
  \BibitemShut {NoStop}%
\bibitem [{\citenamefont {Elvin-Poole}\ \emph {et~al.}(2023)\citenamefont
  {Elvin-Poole}, \citenamefont {MacCrann} \emph {et~al.}}]{DES:magnification}%
  \BibitemOpen
  \bibfield  {author} {\bibinfo {author} {\bibfnamefont {J.}~\bibnamefont
  {Elvin-Poole}}, \bibinfo {author} {\bibfnamefont {N.}~\bibnamefont
  {MacCrann}}, \emph {et~al.} (\bibinfo {collaboration} {DES}),\ }\href
  {https://doi.org/10.1093/mnras/stad1594} {\bibfield  {journal} {\bibinfo
  {journal} {Mon. Not. Roy. Astron. Soc.}\ }\textbf {\bibinfo {volume} {523}},\
  \bibinfo {pages} {3649} (\bibinfo {year} {2023})},\ \Eprint
  {https://arxiv.org/abs/2209.09782} {arXiv:2209.09782 [astro-ph.CO]}
  \BibitemShut {NoStop}%
\bibitem [{\citenamefont {Pandey}\ \emph {et~al.}(2023)\citenamefont {Pandey},
  \citenamefont {S\'anchez},\ and\ \citenamefont {Jain}}]{Pandey:2023tjn}%
  \BibitemOpen
  \bibfield  {author} {\bibinfo {author} {\bibfnamefont {S.}~\bibnamefont
  {Pandey}}, \bibinfo {author} {\bibfnamefont {C.}~\bibnamefont {S\'anchez}},\
  and\ \bibinfo {author} {\bibfnamefont {B.}~\bibnamefont {Jain}} (\bibinfo
  {collaboration} {LSST Dark Energy Science}),\ }\href@noop {} {\  (\bibinfo
  {year} {2023})},\ \Eprint {https://arxiv.org/abs/2310.01315}
  {arXiv:2310.01315 [astro-ph.CO]} \BibitemShut {NoStop}%
\bibitem [{\citenamefont {Schaan}\ \emph {et~al.}(2020)\citenamefont {Schaan},
  \citenamefont {Ferraro},\ and\ \citenamefont {Seljak}}]{Schaan:2020qox}%
  \BibitemOpen
  \bibfield  {author} {\bibinfo {author} {\bibfnamefont {E.}~\bibnamefont
  {Schaan}}, \bibinfo {author} {\bibfnamefont {S.}~\bibnamefont {Ferraro}},\
  and\ \bibinfo {author} {\bibfnamefont {U.}~\bibnamefont {Seljak}},\ }\href
  {https://doi.org/10.1088/1475-7516/2020/12/001} {\bibfield  {journal}
  {\bibinfo  {journal} {JCAP}\ }\textbf {\bibinfo {volume} {12}},\ \bibinfo
  {pages} {001}},\ \Eprint {https://arxiv.org/abs/2007.12795} {arXiv:2007.12795
  [astro-ph.CO]} \BibitemShut {NoStop}%
\bibitem [{\citenamefont {Fang}\ \emph {et~al.}(2023)\citenamefont {Fang},
  \citenamefont {Krause}, \citenamefont {Eifler}, \citenamefont {Ferraro},
  \citenamefont {Benabed}, \citenamefont {S.}, \citenamefont {Ay\c{c}oberry},
  \citenamefont {Dubois},\ and\ \citenamefont {Miranda}}]{Fang:2023efj}%
  \BibitemOpen
  \bibfield  {author} {\bibinfo {author} {\bibfnamefont {X.}~\bibnamefont
  {Fang}}, \bibinfo {author} {\bibfnamefont {E.}~\bibnamefont {Krause}},
  \bibinfo {author} {\bibfnamefont {T.}~\bibnamefont {Eifler}}, \bibinfo
  {author} {\bibfnamefont {S.}~\bibnamefont {Ferraro}}, \bibinfo {author}
  {\bibfnamefont {K.}~\bibnamefont {Benabed}}, \bibinfo {author} {\bibfnamefont
  {P.~R.}\ \bibnamefont {S.}}, \bibinfo {author} {\bibfnamefont
  {E.}~\bibnamefont {Ay\c{c}oberry}}, \bibinfo {author} {\bibfnamefont
  {Y.}~\bibnamefont {Dubois}},\ and\ \bibinfo {author} {\bibfnamefont
  {V.}~\bibnamefont {Miranda}},\ }\href@noop {} {\  (\bibinfo {year} {2023})},\
  \Eprint {https://arxiv.org/abs/2308.01856} {arXiv:2308.01856 [astro-ph.CO]}
  \BibitemShut {NoStop}%
\bibitem [{\citenamefont {Prat}\ \emph {et~al.}(2022)\citenamefont {Prat} \emph
  {et~al.}}]{LSSTPratZuntz:2022sql}%
  \BibitemOpen
  \bibfield  {author} {\bibinfo {author} {\bibfnamefont {J.}~\bibnamefont
  {Prat}} \emph {et~al.} (\bibinfo {collaboration} {LSST})\ }\href
  {https://doi.org/10.21105/astro.2212.09345} {10.21105/astro.2212.09345}
  (\bibinfo {year} {2022}),\ \Eprint {https://arxiv.org/abs/2212.09345}
  {arXiv:2212.09345 [astro-ph.CO]} \BibitemShut {NoStop}%
\bibitem [{\citenamefont {Blanchard}\ \emph {et~al.}(2020)\citenamefont
  {Blanchard} \emph {et~al.}}]{Euclid:forecast_validation}%
  \BibitemOpen
  \bibfield  {author} {\bibinfo {author} {\bibfnamefont {A.}~\bibnamefont
  {Blanchard}} \emph {et~al.} (\bibinfo {collaboration} {Euclid}),\ }\href
  {https://doi.org/10.1051/0004-6361/202038071} {\bibfield  {journal} {\bibinfo
   {journal} {Astron. Astrophys.}\ }\textbf {\bibinfo {volume} {642}},\
  \bibinfo {pages} {A191} (\bibinfo {year} {2020})},\ \Eprint
  {https://arxiv.org/abs/1910.09273} {arXiv:1910.09273 [astro-ph.CO]}
  \BibitemShut {NoStop}%
\bibitem [{\citenamefont {Tanidis}\ \emph {et~al.}(2023)\citenamefont {Tanidis}
  \emph {et~al.}}]{Euclid:rsd}%
  \BibitemOpen
  \bibfield  {author} {\bibinfo {author} {\bibfnamefont {K.}~\bibnamefont
  {Tanidis}} \emph {et~al.} (\bibinfo {collaboration} {Euclid}),\ }\href@noop
  {} {\  (\bibinfo {year} {2023})},\ \Eprint {https://arxiv.org/abs/2309.00052}
  {arXiv:2309.00052 [astro-ph.CO]} \BibitemShut {NoStop}%
\bibitem [{\citenamefont {Tanidis}\ and\ \citenamefont
  {Camera}(2019)}]{TanidisCamera:rsd:2019teo}%
  \BibitemOpen
  \bibfield  {author} {\bibinfo {author} {\bibfnamefont {K.}~\bibnamefont
  {Tanidis}}\ and\ \bibinfo {author} {\bibfnamefont {S.}~\bibnamefont
  {Camera}},\ }\href {https://doi.org/10.1093/mnras/stz2366} {\bibfield
  {journal} {\bibinfo  {journal} {Mon. Not. Roy. Astron. Soc.}\ }\textbf
  {\bibinfo {volume} {489}},\ \bibinfo {pages} {3385} (\bibinfo {year}
  {2019})},\ \Eprint {https://arxiv.org/abs/1902.07226} {arXiv:1902.07226
  [astro-ph.CO]} \BibitemShut {NoStop}%
\bibitem [{\citenamefont {Lorenz}\ \emph {et~al.}(2018)\citenamefont {Lorenz},
  \citenamefont {Alonso},\ and\ \citenamefont {Ferreira}}]{Lorenz:2017iez}%
  \BibitemOpen
  \bibfield  {author} {\bibinfo {author} {\bibfnamefont {C.~S.}\ \bibnamefont
  {Lorenz}}, \bibinfo {author} {\bibfnamefont {D.}~\bibnamefont {Alonso}},\
  and\ \bibinfo {author} {\bibfnamefont {P.~G.}\ \bibnamefont {Ferreira}},\
  }\href {https://doi.org/10.1103/PhysRevD.97.023537} {\bibfield  {journal}
  {\bibinfo  {journal} {Phys. Rev. D}\ }\textbf {\bibinfo {volume} {97}},\
  \bibinfo {pages} {023537} (\bibinfo {year} {2018})},\ \Eprint
  {https://arxiv.org/abs/1710.02477} {arXiv:1710.02477 [astro-ph.CO]}
  \BibitemShut {NoStop}%
\bibitem [{\citenamefont {Mahony}\ \emph {et~al.}(2022)\citenamefont {Mahony}
  \emph {et~al.}}]{Mahony:magnification}%
  \BibitemOpen
  \bibfield  {author} {\bibinfo {author} {\bibfnamefont {C.}~\bibnamefont
  {Mahony}} \emph {et~al.} (\bibinfo {collaboration} {LSST Dark Energy
  Science}),\ }\href {https://doi.org/10.1093/mnras/stac872} {\bibfield
  {journal} {\bibinfo  {journal} {Mon. Not. Roy. Astron. Soc.}\ }\textbf
  {\bibinfo {volume} {513}},\ \bibinfo {pages} {1210} (\bibinfo {year}
  {2022})},\ \Eprint {https://arxiv.org/abs/2112.01545} {arXiv:2112.01545
  [astro-ph.CO]} \BibitemShut {NoStop}%
\bibitem [{\citenamefont {Krause}\ \emph {et~al.}(2017)\citenamefont {Krause},
  \citenamefont {Eifler} \emph {et~al.}}]{DES:Y1methods}%
  \BibitemOpen
  \bibfield  {author} {\bibinfo {author} {\bibfnamefont {E.}~\bibnamefont
  {Krause}}, \bibinfo {author} {\bibfnamefont {T.}~\bibnamefont {Eifler}},
  \emph {et~al.} (\bibinfo {collaboration} {DES}),\ }\href@noop {} {\
  (\bibinfo {year} {2017})},\ \Eprint {https://arxiv.org/abs/1706.09359}
  {arXiv:1706.09359 [astro-ph.CO]} \BibitemShut {NoStop}%
\bibitem [{\citenamefont {Krause}\ \emph {et~al.}(2021)\citenamefont {Krause}
  \emph {et~al.}}]{DES:Y3methods}%
  \BibitemOpen
  \bibfield  {author} {\bibinfo {author} {\bibfnamefont {E.}~\bibnamefont
  {Krause}} \emph {et~al.} (\bibinfo {collaboration} {DES}),\ }\href@noop {} {\
   (\bibinfo {year} {2021})},\ \Eprint {https://arxiv.org/abs/2105.13548}
  {arXiv:2105.13548 [astro-ph.CO]} \BibitemShut {NoStop}%
\bibitem [{\citenamefont {Fang}\ \emph {et~al.}(2020)\citenamefont {Fang},
  \citenamefont {Krause}, \citenamefont {Eifler},\ and\ \citenamefont
  {MacCrann}}]{Fang:2020}%
  \BibitemOpen
  \bibfield  {author} {\bibinfo {author} {\bibfnamefont {X.}~\bibnamefont
  {Fang}}, \bibinfo {author} {\bibfnamefont {E.}~\bibnamefont {Krause}},
  \bibinfo {author} {\bibfnamefont {T.}~\bibnamefont {Eifler}},\ and\ \bibinfo
  {author} {\bibfnamefont {N.}~\bibnamefont {MacCrann}},\ }\href
  {https://doi.org/10.1088/1475-7516/2020/05/010} {\bibfield  {journal}
  {\bibinfo  {journal} {JCAP}\ }\textbf {\bibinfo {volume} {05}},\ \bibinfo
  {pages} {010}},\ \Eprint {https://arxiv.org/abs/1911.11947} {arXiv:1911.11947
  [astro-ph.CO]} \BibitemShut {NoStop}%
\bibitem [{\citenamefont {Cordero}\ \emph {et~al.}(2022)\citenamefont {Cordero}
  \emph {et~al.}}]{DES:hyperrank}%
  \BibitemOpen
  \bibfield  {author} {\bibinfo {author} {\bibfnamefont {J.~P.}\ \bibnamefont
  {Cordero}} \emph {et~al.} (\bibinfo {collaboration} {DES}),\ }\href
  {https://doi.org/10.1093/mnras/stac147} {\bibfield  {journal} {\bibinfo
  {journal} {Mon. Not. Roy. Astron. Soc.}\ }\textbf {\bibinfo {volume} {511}},\
  \bibinfo {pages} {2170} (\bibinfo {year} {2022})},\ \Eprint
  {https://arxiv.org/abs/2109.09636} {arXiv:2109.09636 [astro-ph.CO]}
  \BibitemShut {NoStop}%
\bibitem [{\citenamefont {Hadzhiyska}\ \emph {et~al.}(2020)\citenamefont
  {Hadzhiyska}, \citenamefont {Alonso}, \citenamefont {Nicola},\ and\
  \citenamefont {Slosar}}]{Hadzhiyska:2020xob}%
  \BibitemOpen
  \bibfield  {author} {\bibinfo {author} {\bibfnamefont {B.}~\bibnamefont
  {Hadzhiyska}}, \bibinfo {author} {\bibfnamefont {D.}~\bibnamefont {Alonso}},
  \bibinfo {author} {\bibfnamefont {A.}~\bibnamefont {Nicola}},\ and\ \bibinfo
  {author} {\bibfnamefont {A.}~\bibnamefont {Slosar}},\ }\href
  {https://doi.org/10.1088/1475-7516/2020/10/056} {\bibfield  {journal}
  {\bibinfo  {journal} {JCAP}\ }\textbf {\bibinfo {volume} {10}},\ \bibinfo
  {pages} {056}},\ \Eprint {https://arxiv.org/abs/2007.14989} {arXiv:2007.14989
  [astro-ph.CO]} \BibitemShut {NoStop}%
\bibitem [{\citenamefont {Cawthon}\ \emph {et~al.}(2022)\citenamefont {Cawthon}
  \emph {et~al.}}]{DES:lenswz}%
  \BibitemOpen
  \bibfield  {author} {\bibinfo {author} {\bibfnamefont {R.}~\bibnamefont
  {Cawthon}} \emph {et~al.} (\bibinfo {collaboration} {DES}),\ }\href
  {https://doi.org/10.1093/mnras/stac1160} {\bibfield  {journal} {\bibinfo
  {journal} {Mon. Not. Roy. Astron. Soc.}\ }\textbf {\bibinfo {volume} {513}},\
  \bibinfo {pages} {5517} (\bibinfo {year} {2022})},\ \Eprint
  {https://arxiv.org/abs/2012.12826} {arXiv:2012.12826 [astro-ph.CO]}
  \BibitemShut {NoStop}%
\bibitem [{\citenamefont {Giannini}\ \emph {et~al.}(2024)\citenamefont
  {Giannini} \emph {et~al.}}]{DES:2022redshiftcal}%
  \BibitemOpen
  \bibfield  {author} {\bibinfo {author} {\bibfnamefont {G.}~\bibnamefont
  {Giannini}} \emph {et~al.},\ }\href {https://doi.org/10.1093/mnras/stad2945}
  {\bibfield  {journal} {\bibinfo  {journal} {Mon. Not. Roy. Astron. Soc.}\
  }\textbf {\bibinfo {volume} {527}},\ \bibinfo {pages} {2010} (\bibinfo {year}
  {2024})},\ \Eprint {https://arxiv.org/abs/2209.05853} {arXiv:2209.05853
  [astro-ph.CO]} \BibitemShut {NoStop}%
\bibitem [{\citenamefont {{Kaiser}}(1987)}]{Kaiser1987}%
  \BibitemOpen
  \bibfield  {author} {\bibinfo {author} {\bibfnamefont {N.}~\bibnamefont
  {{Kaiser}}},\ }\href {https://doi.org/10.1093/mnras/227.1.1} {\bibfield
  {journal} {\bibinfo  {journal} {Mon. Not. Roy. Astron. Soc.}\ }\textbf
  {\bibinfo {volume} {227}},\ \bibinfo {pages} {1} (\bibinfo {year}
  {1987})}\BibitemShut {NoStop}%
\bibitem [{\citenamefont {{Lewis}}\ and\ \citenamefont
  {{Challinor}}(2011)}]{CAMB}%
  \BibitemOpen
  \bibfield  {author} {\bibinfo {author} {\bibfnamefont {A.}~\bibnamefont
  {{Lewis}}}\ and\ \bibinfo {author} {\bibfnamefont {A.}~\bibnamefont
  {{Challinor}}},\ }\href@noop {} {\bibinfo {title} {{CAMB: Code for
  Anisotropies in the Microwave Background}}},\ \bibinfo {howpublished}
  {Astrophysics Source Code Library, record ascl:1102.026} (\bibinfo {year}
  {2011}),\ \Eprint {https://arxiv.org/abs/1102.026} {ascl:1102.026}
  \BibitemShut {NoStop}%
\bibitem [{\citenamefont {Lewis}\ \emph {et~al.}(2000)\citenamefont {Lewis},
  \citenamefont {Challinor},\ and\ \citenamefont {Lasenby}}]{Lewis:1999bs}%
  \BibitemOpen
  \bibfield  {author} {\bibinfo {author} {\bibfnamefont {A.}~\bibnamefont
  {Lewis}}, \bibinfo {author} {\bibfnamefont {A.}~\bibnamefont {Challinor}},\
  and\ \bibinfo {author} {\bibfnamefont {A.}~\bibnamefont {Lasenby}},\ }\href
  {https://doi.org/10.1086/309179} {\bibfield  {journal} {\bibinfo  {journal}
  {\apj}\ }\textbf {\bibinfo {volume} {538}},\ \bibinfo {pages} {473} (\bibinfo
  {year} {2000})},\ \Eprint {https://arxiv.org/abs/astro-ph/9911177}
  {arXiv:astro-ph/9911177 [astro-ph]} \BibitemShut {NoStop}%
\bibitem [{\citenamefont {Howlett}\ \emph {et~al.}(2012)\citenamefont
  {Howlett}, \citenamefont {Lewis}, \citenamefont {Hall},\ and\ \citenamefont
  {Challinor}}]{Howlett:2012mh}%
  \BibitemOpen
  \bibfield  {author} {\bibinfo {author} {\bibfnamefont {C.}~\bibnamefont
  {Howlett}}, \bibinfo {author} {\bibfnamefont {A.}~\bibnamefont {Lewis}},
  \bibinfo {author} {\bibfnamefont {A.}~\bibnamefont {Hall}},\ and\ \bibinfo
  {author} {\bibfnamefont {A.}~\bibnamefont {Challinor}},\ }\href
  {https://doi.org/10.1088/1475-7516/2012/04/027} {\bibfield  {journal}
  {\bibinfo  {journal} {JCAP}\ }\textbf {\bibinfo {volume} {1204}},\ \bibinfo
  {pages} {027}},\ \Eprint {https://arxiv.org/abs/1201.3654} {arXiv:1201.3654
  [astro-ph.CO]} \BibitemShut {NoStop}%
\bibitem [{\citenamefont {Challinor}\ and\ \citenamefont
  {Lewis}(2011)}]{Challinor:2011bk}%
  \BibitemOpen
  \bibfield  {author} {\bibinfo {author} {\bibfnamefont {A.}~\bibnamefont
  {Challinor}}\ and\ \bibinfo {author} {\bibfnamefont {A.}~\bibnamefont
  {Lewis}},\ }\href {https://doi.org/10.1103/PhysRevD.84.043516} {\bibfield
  {journal} {\bibinfo  {journal} {\prd}\ }\textbf {\bibinfo {volume} {84}},\
  \bibinfo {pages} {043516} (\bibinfo {year} {2011})}\BibitemShut {NoStop}%
\bibitem [{\citenamefont {De~Vicente}\ \emph {et~al.}(2016)\citenamefont
  {De~Vicente}, \citenamefont {S\'anchez},\ and\ \citenamefont
  {Sevilla-Noarbe}}]{DeVicente:2015kyp}%
  \BibitemOpen
  \bibfield  {author} {\bibinfo {author} {\bibfnamefont {J.}~\bibnamefont
  {De~Vicente}}, \bibinfo {author} {\bibfnamefont {E.}~\bibnamefont
  {S\'anchez}},\ and\ \bibinfo {author} {\bibfnamefont {I.}~\bibnamefont
  {Sevilla-Noarbe}},\ }\href {https://doi.org/10.1093/mnras/stw857} {\bibfield
  {journal} {\bibinfo  {journal} {Mon. Not. Roy. Astron. Soc.}\ }\textbf
  {\bibinfo {volume} {459}},\ \bibinfo {pages} {3078} (\bibinfo {year}
  {2016})},\ \Eprint {https://arxiv.org/abs/1511.07623} {arXiv:1511.07623
  [astro-ph.CO]} \BibitemShut {NoStop}%
\bibitem [{\citenamefont {Sevilla-Noarbe}\ \emph {et~al.}(2021)\citenamefont
  {Sevilla-Noarbe} \emph {et~al.}}]{DES:2020aks}%
  \BibitemOpen
  \bibfield  {author} {\bibinfo {author} {\bibfnamefont {I.}~\bibnamefont
  {Sevilla-Noarbe}} \emph {et~al.} (\bibinfo {collaboration} {DES}),\ }\href
  {https://doi.org/10.3847/1538-4365/abeb66} {\bibfield  {journal} {\bibinfo
  {journal} {Astrophys. J. Suppl.}\ }\textbf {\bibinfo {volume} {254}},\
  \bibinfo {pages} {24} (\bibinfo {year} {2021})},\ \Eprint
  {https://arxiv.org/abs/2011.03407} {arXiv:2011.03407 [astro-ph.CO]}
  \BibitemShut {NoStop}%
\bibitem [{\citenamefont {Zhang}\ \emph {et~al.}(2022)\citenamefont {Zhang},
  \citenamefont {Chang}, \citenamefont {Larsen}, \citenamefont {Secco},\ and\
  \citenamefont {Zuntz}}]{Zhang:2022}%
  \BibitemOpen
  \bibfield  {author} {\bibinfo {author} {\bibfnamefont {Z.~J.}\ \bibnamefont
  {Zhang}}, \bibinfo {author} {\bibfnamefont {C.}~\bibnamefont {Chang}},
  \bibinfo {author} {\bibfnamefont {P.}~\bibnamefont {Larsen}}, \bibinfo
  {author} {\bibfnamefont {L.~F.}\ \bibnamefont {Secco}},\ and\ \bibinfo
  {author} {\bibfnamefont {J.}~\bibnamefont {Zuntz}} (\bibinfo {collaboration}
  {LSST Dark Energy Science}),\ }\href {https://doi.org/10.1093/mnras/stac1407}
  {\bibfield  {journal} {\bibinfo  {journal} {Mon. Not. Roy. Astron. Soc.}\
  }\textbf {\bibinfo {volume} {514}},\ \bibinfo {pages} {2181} (\bibinfo {year}
  {2022})},\ \Eprint {https://arxiv.org/abs/2111.04917} {arXiv:2111.04917
  [astro-ph.CO]} \BibitemShut {NoStop}%
\bibitem [{\citenamefont {Zuntz}\ \emph {et~al.}(2015)\citenamefont {Zuntz},
  \citenamefont {Paterno}, \citenamefont {Jennings}, \citenamefont {Rudd},
  \citenamefont {Manzotti}, \citenamefont {Dodelson}, \citenamefont {Bridle},
  \citenamefont {Sehrish},\ and\ \citenamefont {Kowalkowski}}]{cosmosis}%
  \BibitemOpen
  \bibfield  {author} {\bibinfo {author} {\bibfnamefont {J.}~\bibnamefont
  {Zuntz}}, \bibinfo {author} {\bibfnamefont {M.}~\bibnamefont {Paterno}},
  \bibinfo {author} {\bibfnamefont {E.}~\bibnamefont {Jennings}}, \bibinfo
  {author} {\bibfnamefont {D.}~\bibnamefont {Rudd}}, \bibinfo {author}
  {\bibfnamefont {A.}~\bibnamefont {Manzotti}}, \bibinfo {author}
  {\bibfnamefont {S.}~\bibnamefont {Dodelson}}, \bibinfo {author}
  {\bibfnamefont {S.}~\bibnamefont {Bridle}}, \bibinfo {author} {\bibfnamefont
  {S.}~\bibnamefont {Sehrish}},\ and\ \bibinfo {author} {\bibfnamefont
  {J.}~\bibnamefont {Kowalkowski}},\ }\href
  {https://doi.org/10.1016/j.ascom.2015.05.005} {\bibfield  {journal} {\bibinfo
   {journal} {Astron. Comput.}\ }\textbf {\bibinfo {volume} {12}},\ \bibinfo
  {pages} {45} (\bibinfo {year} {2015})},\ \Eprint
  {https://arxiv.org/abs/1409.3409} {arXiv:1409.3409 [astro-ph.CO]}
  \BibitemShut {NoStop}%
\bibitem [{\citenamefont {Verde}(2010)}]{StatMethods}%
  \BibitemOpen
  \bibfield  {author} {\bibinfo {author} {\bibfnamefont {L.}~\bibnamefont
  {Verde}},\ }in\ \href@noop {} {\emph {\bibinfo {booktitle} {Lectures on
  Cosmology}}}\ (\bibinfo  {publisher} {Springer},\ \bibinfo {year} {2010})\
  pp.\ \bibinfo {pages} {147--177}\BibitemShut {NoStop}%
\bibitem [{\citenamefont {Huterer}(2020)}]{MichiganLecNote}%
  \BibitemOpen
  \bibfield  {author} {\bibinfo {author} {\bibfnamefont {D.}~\bibnamefont
  {Huterer}},\ }\href
  {https://sites.google.com/a/umich.edu/cosmology-summer-school-2020/schedule-of-events}
  {\bibinfo {title} {Structure formation in the universe}},\ \bibinfo
  {howpublished} {Lecture Notes, Michigan Cosmology Summer School} (\bibinfo
  {year} {2020})\BibitemShut {NoStop}%
\bibitem [{\citenamefont {Alonso}\ \emph {et~al.}(2015)\citenamefont {Alonso},
  \citenamefont {Bull}, \citenamefont {Ferreira}, \citenamefont {Maartens},\
  and\ \citenamefont {Santos}}]{Alonso_2015}%
  \BibitemOpen
  \bibfield  {author} {\bibinfo {author} {\bibfnamefont {D.}~\bibnamefont
  {Alonso}}, \bibinfo {author} {\bibfnamefont {P.}~\bibnamefont {Bull}},
  \bibinfo {author} {\bibfnamefont {P.~G.}\ \bibnamefont {Ferreira}}, \bibinfo
  {author} {\bibfnamefont {R.}~\bibnamefont {Maartens}},\ and\ \bibinfo
  {author} {\bibfnamefont {M.~G.}\ \bibnamefont {Santos}},\ }\href
  {https://doi.org/10.1088/0004-637x/814/2/145} {\bibfield  {journal} {\bibinfo
   {journal} {The Astrophysical Journal}\ }\textbf {\bibinfo {volume} {814}},\
  \bibinfo {pages} {145} (\bibinfo {year} {2015})}\BibitemShut {NoStop}%
\bibitem [{\citenamefont {{DES Collaboration}}(2023)}]{DES:Y3ext}%
  \BibitemOpen
  \bibfield  {author} {\bibinfo {author} {\bibnamefont {{DES Collaboration}}}
  (\bibinfo {collaboration} {DES}),\ }\href
  {https://doi.org/10.1103/PhysRevD.107.083504} {\bibfield  {journal} {\bibinfo
   {journal} {Phys. Rev. D}\ }\textbf {\bibinfo {volume} {107}},\ \bibinfo
  {pages} {083504} (\bibinfo {year} {2023})},\ \Eprint
  {https://arxiv.org/abs/2207.05766} {arXiv:2207.05766 [astro-ph.CO]}
  \BibitemShut {NoStop}%
\bibitem [{\citenamefont {Tanidis}\ \emph {et~al.}(2020)\citenamefont
  {Tanidis}, \citenamefont {Camera},\ and\ \citenamefont
  {Parkinson}}]{Tanidis:mag:2019fdh}%
  \BibitemOpen
  \bibfield  {author} {\bibinfo {author} {\bibfnamefont {K.}~\bibnamefont
  {Tanidis}}, \bibinfo {author} {\bibfnamefont {S.}~\bibnamefont {Camera}},\
  and\ \bibinfo {author} {\bibfnamefont {D.}~\bibnamefont {Parkinson}},\ }\href
  {https://doi.org/10.1093/mnras/stz3394} {\bibfield  {journal} {\bibinfo
  {journal} {Mon. Not. Roy. Astron. Soc.}\ }\textbf {\bibinfo {volume} {491}},\
  \bibinfo {pages} {4869} (\bibinfo {year} {2020})},\ \Eprint
  {https://arxiv.org/abs/1909.10539} {arXiv:1909.10539 [astro-ph.CO]}
  \BibitemShut {NoStop}%
\bibitem [{\citenamefont {Farren}\ \emph {et~al.}(2023)\citenamefont {Farren}
  \emph {et~al.}}]{ACTFarren:2023oei}%
  \BibitemOpen
  \bibfield  {author} {\bibinfo {author} {\bibfnamefont {G.~S.}\ \bibnamefont
  {Farren}} \emph {et~al.} (\bibinfo {collaboration} {ACT}),\ }\href@noop {} {\
   (\bibinfo {year} {2023})},\ \Eprint {https://arxiv.org/abs/2309.05659}
  {arXiv:2309.05659 [astro-ph.CO]} \BibitemShut {NoStop}%
\bibitem [{\citenamefont {Krolewski}\ \emph {et~al.}(2021)\citenamefont
  {Krolewski}, \citenamefont {Ferraro},\ and\ \citenamefont
  {White}}]{Krolewski:2021yqy}%
  \BibitemOpen
  \bibfield  {author} {\bibinfo {author} {\bibfnamefont {A.}~\bibnamefont
  {Krolewski}}, \bibinfo {author} {\bibfnamefont {S.}~\bibnamefont {Ferraro}},\
  and\ \bibinfo {author} {\bibfnamefont {M.}~\bibnamefont {White}},\ }\href
  {https://doi.org/10.1088/1475-7516/2021/12/028} {\bibfield  {journal}
  {\bibinfo  {journal} {JCAP}\ }\textbf {\bibinfo {volume} {12}}\bibfield
  {number} {\bibinfo  {number} { (12)},\ \bibinfo {pages} {028}},\ }\Eprint
  {https://arxiv.org/abs/2105.03421} {arXiv:2105.03421 [astro-ph.CO]}
  \BibitemShut {NoStop}%
\bibitem [{\citenamefont {Ishak}\ \emph {et~al.}(2019)\citenamefont {Ishak}
  \emph {et~al.}}]{Ishak:2019aay}%
  \BibitemOpen
  \bibfield  {author} {\bibinfo {author} {\bibfnamefont {M.}~\bibnamefont
  {Ishak}} \emph {et~al.},\ }\href@noop {} {\  (\bibinfo {year} {2019})},\
  \Eprint {https://arxiv.org/abs/1905.09687} {arXiv:1905.09687 [astro-ph.CO]}
  \BibitemShut {NoStop}%
\bibitem [{\citenamefont {Garc\'\i{}a-Garc\'\i{}a}\ \emph
  {et~al.}(2021)\citenamefont {Garc\'\i{}a-Garc\'\i{}a}, \citenamefont
  {Zapatero}, \citenamefont {Alonso}, \citenamefont {Bellini}, \citenamefont
  {Ferreira}, \citenamefont {Mueller}, \citenamefont {Nicola},\ and\
  \citenamefont {Ruiz-Lapuente}}]{Garcia-Garcia:2021unp}%
  \BibitemOpen
  \bibfield  {author} {\bibinfo {author} {\bibfnamefont {C.}~\bibnamefont
  {Garc\'\i{}a-Garc\'\i{}a}}, \bibinfo {author} {\bibfnamefont {J.~R.}\
  \bibnamefont {Zapatero}}, \bibinfo {author} {\bibfnamefont {D.}~\bibnamefont
  {Alonso}}, \bibinfo {author} {\bibfnamefont {E.}~\bibnamefont {Bellini}},
  \bibinfo {author} {\bibfnamefont {P.~G.}\ \bibnamefont {Ferreira}}, \bibinfo
  {author} {\bibfnamefont {E.-M.}\ \bibnamefont {Mueller}}, \bibinfo {author}
  {\bibfnamefont {A.}~\bibnamefont {Nicola}},\ and\ \bibinfo {author}
  {\bibfnamefont {P.}~\bibnamefont {Ruiz-Lapuente}},\ }\href
  {https://doi.org/10.1088/1475-7516/2021/10/030} {\bibfield  {journal}
  {\bibinfo  {journal} {JCAP}\ }\textbf {\bibinfo {volume} {10}},\ \bibinfo
  {pages} {030}},\ \Eprint {https://arxiv.org/abs/2105.12108} {arXiv:2105.12108
  [astro-ph.CO]} \BibitemShut {NoStop}%
\bibitem [{\citenamefont {Harris}\ \emph {et~al.}(2020)\citenamefont {Harris},
  \citenamefont {Millman}, \citenamefont {van~der Walt}, \citenamefont
  {Gommers}, \citenamefont {Virtanen}, \citenamefont {Cournapeau},
  \citenamefont {Wieser}, \citenamefont {Taylor}, \citenamefont {Berg},
  \citenamefont {Smith}, \citenamefont {Kern}, \citenamefont {Picus},
  \citenamefont {Hoyer}, \citenamefont {van Kerkwijk}, \citenamefont {Brett},
  \citenamefont {Haldane}, \citenamefont {del R{\'{i}}o}, \citenamefont
  {Wiebe}, \citenamefont {Peterson}, \citenamefont {G{\'{e}}rard-Marchant},
  \citenamefont {Sheppard}, \citenamefont {Reddy}, \citenamefont {Weckesser},
  \citenamefont {Abbasi}, \citenamefont {Gohlke},\ and\ \citenamefont
  {Oliphant}}]{numpy}%
  \BibitemOpen
  \bibfield  {author} {\bibinfo {author} {\bibfnamefont {C.~R.}\ \bibnamefont
  {Harris}}, \bibinfo {author} {\bibfnamefont {K.~J.}\ \bibnamefont {Millman}},
  \bibinfo {author} {\bibfnamefont {S.~J.}\ \bibnamefont {van~der Walt}},
  \bibinfo {author} {\bibfnamefont {R.}~\bibnamefont {Gommers}}, \bibinfo
  {author} {\bibfnamefont {P.}~\bibnamefont {Virtanen}}, \bibinfo {author}
  {\bibfnamefont {D.}~\bibnamefont {Cournapeau}}, \bibinfo {author}
  {\bibfnamefont {E.}~\bibnamefont {Wieser}}, \bibinfo {author} {\bibfnamefont
  {J.}~\bibnamefont {Taylor}}, \bibinfo {author} {\bibfnamefont
  {S.}~\bibnamefont {Berg}}, \bibinfo {author} {\bibfnamefont {N.~J.}\
  \bibnamefont {Smith}}, \bibinfo {author} {\bibfnamefont {R.}~\bibnamefont
  {Kern}}, \bibinfo {author} {\bibfnamefont {M.}~\bibnamefont {Picus}},
  \bibinfo {author} {\bibfnamefont {S.}~\bibnamefont {Hoyer}}, \bibinfo
  {author} {\bibfnamefont {M.~H.}\ \bibnamefont {van Kerkwijk}}, \bibinfo
  {author} {\bibfnamefont {M.}~\bibnamefont {Brett}}, \bibinfo {author}
  {\bibfnamefont {A.}~\bibnamefont {Haldane}}, \bibinfo {author} {\bibfnamefont
  {J.~F.}\ \bibnamefont {del R{\'{i}}o}}, \bibinfo {author} {\bibfnamefont
  {M.}~\bibnamefont {Wiebe}}, \bibinfo {author} {\bibfnamefont
  {P.}~\bibnamefont {Peterson}}, \bibinfo {author} {\bibfnamefont
  {P.}~\bibnamefont {G{\'{e}}rard-Marchant}}, \bibinfo {author} {\bibfnamefont
  {K.}~\bibnamefont {Sheppard}}, \bibinfo {author} {\bibfnamefont
  {T.}~\bibnamefont {Reddy}}, \bibinfo {author} {\bibfnamefont
  {W.}~\bibnamefont {Weckesser}}, \bibinfo {author} {\bibfnamefont
  {H.}~\bibnamefont {Abbasi}}, \bibinfo {author} {\bibfnamefont
  {C.}~\bibnamefont {Gohlke}},\ and\ \bibinfo {author} {\bibfnamefont {T.~E.}\
  \bibnamefont {Oliphant}},\ }\href {https://doi.org/10.1038/s41586-020-2649-2}
  {\bibfield  {journal} {\bibinfo  {journal} {Nature}\ }\textbf {\bibinfo
  {volume} {585}},\ \bibinfo {pages} {357} (\bibinfo {year}
  {2020})}\BibitemShut {NoStop}%
\bibitem [{\citenamefont {Hunter}(2007)}]{matplotlib}%
  \BibitemOpen
  \bibfield  {author} {\bibinfo {author} {\bibfnamefont {J.~D.}\ \bibnamefont
  {Hunter}},\ }\href {https://doi.org/10.1109/MCSE.2007.55} {\bibfield
  {journal} {\bibinfo  {journal} {Computing in Science \& Engineering}\
  }\textbf {\bibinfo {volume} {9}},\ \bibinfo {pages} {90} (\bibinfo {year}
  {2007})}\BibitemShut {NoStop}%
\bibitem [{\citenamefont {Virtanen}\ \emph {et~al.}(2020)\citenamefont
  {Virtanen}, \citenamefont {Gommers}, \citenamefont {Oliphant}, \citenamefont
  {Haberland}, \citenamefont {Reddy}, \citenamefont {Cournapeau}, \citenamefont
  {Burovski}, \citenamefont {Peterson}, \citenamefont {Weckesser},
  \citenamefont {Bright}, \citenamefont {{van der Walt}}, \citenamefont
  {Brett}, \citenamefont {Wilson}, \citenamefont {Millman}, \citenamefont
  {Mayorov}, \citenamefont {Nelson}, \citenamefont {Jones}, \citenamefont
  {Kern}, \citenamefont {Larson}, \citenamefont {Carey}, \citenamefont {Polat},
  \citenamefont {Feng}, \citenamefont {Moore}, \citenamefont {{VanderPlas}},
  \citenamefont {Laxalde}, \citenamefont {Perktold}, \citenamefont {Cimrman},
  \citenamefont {Henriksen}, \citenamefont {Quintero}, \citenamefont {Harris},
  \citenamefont {Archibald}, \citenamefont {Ribeiro}, \citenamefont
  {Pedregosa}, \citenamefont {{van Mulbregt}},\ and\ \citenamefont {{SciPy 1.0
  Contributors}}}]{Scipy}%
  \BibitemOpen
  \bibfield  {author} {\bibinfo {author} {\bibfnamefont {P.}~\bibnamefont
  {Virtanen}}, \bibinfo {author} {\bibfnamefont {R.}~\bibnamefont {Gommers}},
  \bibinfo {author} {\bibfnamefont {T.~E.}\ \bibnamefont {Oliphant}}, \bibinfo
  {author} {\bibfnamefont {M.}~\bibnamefont {Haberland}}, \bibinfo {author}
  {\bibfnamefont {T.}~\bibnamefont {Reddy}}, \bibinfo {author} {\bibfnamefont
  {D.}~\bibnamefont {Cournapeau}}, \bibinfo {author} {\bibfnamefont
  {E.}~\bibnamefont {Burovski}}, \bibinfo {author} {\bibfnamefont
  {P.}~\bibnamefont {Peterson}}, \bibinfo {author} {\bibfnamefont
  {W.}~\bibnamefont {Weckesser}}, \bibinfo {author} {\bibfnamefont
  {J.}~\bibnamefont {Bright}}, \bibinfo {author} {\bibfnamefont {S.~J.}\
  \bibnamefont {{van der Walt}}}, \bibinfo {author} {\bibfnamefont
  {M.}~\bibnamefont {Brett}}, \bibinfo {author} {\bibfnamefont
  {J.}~\bibnamefont {Wilson}}, \bibinfo {author} {\bibfnamefont {K.~J.}\
  \bibnamefont {Millman}}, \bibinfo {author} {\bibfnamefont {N.}~\bibnamefont
  {Mayorov}}, \bibinfo {author} {\bibfnamefont {A.~R.~J.}\ \bibnamefont
  {Nelson}}, \bibinfo {author} {\bibfnamefont {E.}~\bibnamefont {Jones}},
  \bibinfo {author} {\bibfnamefont {R.}~\bibnamefont {Kern}}, \bibinfo {author}
  {\bibfnamefont {E.}~\bibnamefont {Larson}}, \bibinfo {author} {\bibfnamefont
  {C.~J.}\ \bibnamefont {Carey}}, \bibinfo {author} {\bibfnamefont
  {{\.I}.}~\bibnamefont {Polat}}, \bibinfo {author} {\bibfnamefont
  {Y.}~\bibnamefont {Feng}}, \bibinfo {author} {\bibfnamefont {E.~W.}\
  \bibnamefont {Moore}}, \bibinfo {author} {\bibfnamefont {J.}~\bibnamefont
  {{VanderPlas}}}, \bibinfo {author} {\bibfnamefont {D.}~\bibnamefont
  {Laxalde}}, \bibinfo {author} {\bibfnamefont {J.}~\bibnamefont {Perktold}},
  \bibinfo {author} {\bibfnamefont {R.}~\bibnamefont {Cimrman}}, \bibinfo
  {author} {\bibfnamefont {I.}~\bibnamefont {Henriksen}}, \bibinfo {author}
  {\bibfnamefont {E.~A.}\ \bibnamefont {Quintero}}, \bibinfo {author}
  {\bibfnamefont {C.~R.}\ \bibnamefont {Harris}}, \bibinfo {author}
  {\bibfnamefont {A.~M.}\ \bibnamefont {Archibald}}, \bibinfo {author}
  {\bibfnamefont {A.~H.}\ \bibnamefont {Ribeiro}}, \bibinfo {author}
  {\bibfnamefont {F.}~\bibnamefont {Pedregosa}}, \bibinfo {author}
  {\bibfnamefont {P.}~\bibnamefont {{van Mulbregt}}},\ and\ \bibinfo {author}
  {\bibnamefont {{SciPy 1.0 Contributors}}},\ }\href
  {https://doi.org/10.1038/s41592-019-0686-2} {\bibfield  {journal} {\bibinfo
  {journal} {Nature Methods}\ }\textbf {\bibinfo {volume} {17}},\ \bibinfo
  {pages} {261} (\bibinfo {year} {2020})}\BibitemShut {NoStop}%
\bibitem [{\citenamefont {Carron}(2013)}]{Carron:2013}%
  \BibitemOpen
  \bibfield  {author} {\bibinfo {author} {\bibfnamefont {J.}~\bibnamefont
  {Carron}},\ }\href {https://doi.org/10.1051/0004-6361/201220538} {\bibfield
  {journal} {\bibinfo  {journal} {Astronomy \& Astrophysics}\ }\textbf
  {\bibinfo {volume} {551}},\ \bibinfo {pages} {A88} (\bibinfo {year}
  {2013})}\BibitemShut {NoStop}%
\end{thebibliography}%

\end{document}